\documentclass[12pt]{article}
\usepackage{sectsty}
\usepackage{bm}
\usepackage[capposition=top]{floatrow}

\usepackage[T1]{fontenc}
\usepackage[utf8]{inputenc}
\usepackage{fourier}
\usepackage{natbib}

\newcommand{\hide}[1]{}
\usepackage{authblk}
\usepackage{multirow}
\usepackage{pdflscape}
\usepackage{geometry}
\usepackage{placeins, setspace}
\usepackage{ragged2e}
\usepackage{pdflscape, chngpage}
\usepackage{url}
\usepackage[plainpages=false]{hyperref}
\hypersetup{colorlinks=true, urlcolor=Cerulean, citecolor=Cerulean, linkcolor=Cerulean}
\usepackage[usenames, dvipsnames, svgnames, table]{xcolor}
\usepackage{pdfpages}

\usepackage{booktabs, dcolumn, longtable}

\usepackage{caption, subcaption}
\usepackage{graphicx}
\usepackage{pst-all}
\usepackage{cprotect}

\usepackage{amsmath, amssymb, amsthm}
\usepackage{nicefrac}
\usepackage{float}
\usepackage{mathtools}

\usepackage{listings}
\usepackage{fancyvrb}
\VerbatimFootnotes

\definecolor{snow}{rgb}{1.000,0.980,0.980}
\definecolor{snow1}{rgb}{1.000,0.980,0.980}
\definecolor{snow2}{rgb}{0.933,0.914,0.914}
\definecolor{snow3}{rgb}{0.804,0.788,0.788}
\definecolor{snow4}{rgb}{0.545,0.537,0.537}
\definecolor{GhostWhite}{rgb}{0.973,0.973,1.000}
\definecolor{WhiteSmoke}{rgb}{0.961,0.961,0.961}
\definecolor{gainsboro}{rgb}{0.863,0.863,0.863}
\definecolor{FloralWhite}{rgb}{1.000,0.980,0.941}
\definecolor{OldLace}{rgb}{0.992,0.961,0.902}
\definecolor{linen}{rgb}{0.980,0.941,0.902}
\definecolor{AntiqueWhite}{rgb}{0.980,0.922,0.843}
\definecolor{PapayaWhip}{rgb}{1.000,0.937,0.835}
\definecolor{BlanchedAlmond}{rgb}{1.000,0.922,0.804}
\definecolor{bisque}{rgb}{1.000,0.894,0.769}
\definecolor{PeachPuff}{rgb}{1.000,0.855,0.725}
\definecolor{NavajoWhite}{rgb}{1.000,0.871,0.678}
\definecolor{moccasin}{rgb}{1.000,0.894,0.710}
\definecolor{cornsilk}{rgb}{1.000,0.973,0.863}
\definecolor{ivory}{rgb}{1.000,1.000,0.941}
\definecolor{LemonChiffon}{rgb}{1.000,0.980,0.804}
\definecolor{seashell}{rgb}{1.000,0.961,0.933}
\definecolor{honeydew}{rgb}{0.941,1.000,0.941}
\definecolor{MintCream}{rgb}{0.961,1.000,0.980}
\definecolor{azure}{rgb}{0.941,1.000,1.000}
\definecolor{AliceBlue}{rgb}{0.941,0.973,1.000}
\definecolor{lavender}{rgb}{0.902,0.902,0.980}
\definecolor{LavenderBlush}{rgb}{1.000,0.941,0.961}
\definecolor{MistyRose}{rgb}{1.000,0.894,0.882}
\definecolor{white}{rgb}{1.000,1.000,1.000}
\definecolor{black}{rgb}{0.000,0.000,0.000}
\definecolor{DarkSlateGray}{rgb}{0.184,0.310,0.310}
\definecolor{DimGray}{rgb}{0.412,0.412,0.412}
\definecolor{SlateGray}{rgb}{0.439,0.502,0.565}
\definecolor{LightSlateGray}{rgb}{0.467,0.533,0.600}
\definecolor{gray}{rgb}{0.745,0.745,0.745}
\definecolor{LightGray}{rgb}{0.827,0.827,0.827}
\definecolor{MidnightBlue}{rgb}{0.098,0.098,0.439}
\definecolor{EnumerateBlue}{rgb}{0.2,0.2,0.701}
\definecolor{NavyBlue}{rgb}{0.000,0.000,0.502}
\definecolor{CornflowerBlue}{rgb}{0.392,0.584,0.929}
\definecolor{DarkSlateBlue}{rgb}{0.282,0.239,0.545}
\definecolor{SlateBlue}{rgb}{0.416,0.353,0.804}
\definecolor{MediumSlateBlue}{rgb}{0.482,0.408,0.933}
\definecolor{LightSlateBlue}{rgb}{0.518,0.439,1.000}
\definecolor{MediumBlue}{rgb}{0.000,0.000,0.804}
\definecolor{RoyalBlue}{rgb}{0.255,0.412,0.882}
\definecolor{blue}{rgb}{0.000,0.000,1.000}
\definecolor{DodgerBlue}{rgb}{0.118,0.565,1.000}
\definecolor{DeepSkyBlue}{rgb}{0.000,0.749,1.000}
\definecolor{SkyBlue}{rgb}{0.529,0.808,0.922}
\definecolor{LightSkyBlue}{rgb}{0.529,0.808,0.980}
\definecolor{SteelBlue}{rgb}{0.275,0.510,0.706}
\definecolor{LightSteelBlue}{rgb}{0.690,0.769,0.871}
\definecolor{LightBlue}{rgb}{0.678,0.847,0.902}
\definecolor{PowderBlue}{rgb}{0.690,0.878,0.902}
\definecolor{PaleTurquoise}{rgb}{0.686,0.933,0.933}
\definecolor{DarkTurquoise}{rgb}{0.000,0.808,0.820}
\definecolor{MediumTurquoise}{rgb}{0.282,0.820,0.800}
\definecolor{turquoise}{rgb}{0.251,0.878,0.816}
\definecolor{cyan}{rgb}{0.000,1.000,1.000}
\definecolor{LightCyan}{rgb}{0.878,1.000,1.000}
\definecolor{CadetBlue}{rgb}{0.373,0.620,0.627}
\definecolor{MediumAquamarine}{rgb}{0.400,0.804,0.667}
\definecolor{aquamarine}{rgb}{0.498,1.000,0.831}
\definecolor{DarkGreen}{rgb}{0.000,0.392,0.000}
\definecolor{DarkOliveGreen}{rgb}{0.333,0.420,0.184}
\definecolor{DarkSeaGreen}{rgb}{0.561,0.737,0.561}
\definecolor{SeaGreen}{rgb}{0.180,0.545,0.341}
\definecolor{MediumSeaGreen}{rgb}{0.235,0.702,0.443}
\definecolor{LightSeaGreen}{rgb}{0.125,0.698,0.667}
\definecolor{PaleGreen}{rgb}{0.596,0.984,0.596}
\definecolor{SpringGreen}{rgb}{0.000,1.000,0.498}
\definecolor{LawnGreen}{rgb}{0.486,0.988,0.000}
\definecolor{green}{rgb}{0.000,1.000,0.000}
\definecolor{chartreuse}{rgb}{0.498,1.000,0.000}
\definecolor{MediumSpringGreen}{rgb}{0.000,0.980,0.604}
\definecolor{GreenYellow}{rgb}{0.678,1.000,0.184}
\definecolor{LimeGreen}{rgb}{0.196,0.804,0.196}
\definecolor{YellowGreen}{rgb}{0.604,0.804,0.196}
\definecolor{ForestGreen}{rgb}{0.133,0.545,0.133}
\definecolor{OliveDrab}{rgb}{0.420,0.557,0.137}
\definecolor{DarkKhaki}{rgb}{0.741,0.718,0.420}
\definecolor{khaki}{rgb}{0.941,0.902,0.549}
\definecolor{PaleGoldenrod}{rgb}{0.933,0.910,0.667}
\definecolor{LightGoldenrodYellow}{rgb}{0.980,0.980,0.824}
\definecolor{LightYellow}{rgb}{1.000,1.000,0.878}
\definecolor{yellow}{rgb}{1.000,1.000,0.000}
\definecolor{gold}{rgb}{1.000,0.843,0.000}
\definecolor{LightGoldenrod}{rgb}{0.933,0.867,0.510}
\definecolor{goldenrod}{rgb}{0.855,0.647,0.125}
\definecolor{DarkGoldenrod}{rgb}{0.722,0.525,0.043}
\definecolor{RosyBrown}{rgb}{0.737,0.561,0.561}
\definecolor{IndianRed}{rgb}{0.804,0.361,0.361}
\definecolor{SaddleBrown}{rgb}{0.545,0.271,0.075}
\definecolor{sienna}{rgb}{0.627,0.322,0.176}
\definecolor{peru}{rgb}{0.804,0.522,0.247}
\definecolor{burlywood}{rgb}{0.871,0.722,0.529}
\definecolor{beige}{rgb}{0.961,0.961,0.863}
\definecolor{wheat}{rgb}{0.961,0.871,0.702}
\definecolor{SandyBrown}{rgb}{0.957,0.643,0.376}
\definecolor{tan}{rgb}{0.824,0.706,0.549}
\definecolor{chocolate}{rgb}{0.824,0.412,0.118}
\definecolor{firebrick}{rgb}{0.698,0.133,0.133}
\definecolor{brown}{rgb}{0.647,0.165,0.165}
\definecolor{DarkSalmon}{rgb}{0.914,0.588,0.478}
\definecolor{salmon}{rgb}{0.980,0.502,0.447}
\definecolor{LightSalmon}{rgb}{1.000,0.627,0.478}
\definecolor{orange}{rgb}{1.000,0.647,0.000}
\definecolor{DarkOrange}{rgb}{1.000,0.549,0.000}
\definecolor{coral}{rgb}{1.000,0.498,0.314}
\definecolor{LightCoral}{rgb}{0.941,0.502,0.502}
\definecolor{tomato}{rgb}{1.000,0.388,0.278}
\definecolor{OrangeRed}{rgb}{1.000,0.271,0.000}
\definecolor{red}{rgb}{1.000,0.000,0.000}
\definecolor{HotPink}{rgb}{1.000,0.412,0.706}
\definecolor{DeepPink}{rgb}{1.000,0.078,0.576}
\definecolor{pink}{rgb}{1.000,0.753,0.796}
\definecolor{LightPink}{rgb}{1.000,0.714,0.757}
\definecolor{PaleVioletRed}{rgb}{0.859,0.439,0.576}
\definecolor{maroon}{rgb}{0.690,0.188,0.376}
\definecolor{MediumVioletRed}{rgb}{0.780,0.082,0.522}
\definecolor{VioletRed}{rgb}{0.816,0.125,0.565}
\definecolor{magenta}{rgb}{1.000,0.000,1.000}
\definecolor{violet}{rgb}{0.933,0.510,0.933}
\definecolor{plum}{rgb}{0.867,0.627,0.867}
\definecolor{orchid}{rgb}{0.855,0.439,0.839}
\definecolor{MediumOrchid}{rgb}{0.729,0.333,0.827}
\definecolor{DarkOrchid}{rgb}{0.600,0.196,0.800}
\definecolor{DarkViolet}{rgb}{0.580,0.000,0.827}
\definecolor{BlueViolet}{rgb}{0.541,0.169,0.886}
\definecolor{purple}{rgb}{0.627,0.125,0.941}
\definecolor{MediumPurple}{rgb}{0.576,0.439,0.859}
\definecolor{thistle}{rgb}{0.847,0.749,0.847}
\definecolor{seashell1}{rgb}{1.000,0.961,0.933}
\definecolor{seashell2}{rgb}{0.933,0.898,0.871}
\definecolor{seashell3}{rgb}{0.804,0.773,0.749}
\definecolor{seashell4}{rgb}{0.545,0.525,0.510}
\definecolor{AntiqueWhite1}{rgb}{1.000,0.937,0.859}
\definecolor{AntiqueWhite2}{rgb}{0.933,0.875,0.800}
\definecolor{AntiqueWhite3}{rgb}{0.804,0.753,0.690}
\definecolor{AntiqueWhite4}{rgb}{0.545,0.514,0.471}
\definecolor{bisque1}{rgb}{1.000,0.894,0.769}
\definecolor{bisque2}{rgb}{0.933,0.835,0.718}
\definecolor{bisque3}{rgb}{0.804,0.718,0.620}
\definecolor{bisque4}{rgb}{0.545,0.490,0.420}
\definecolor{PeachPuff1}{rgb}{1.000,0.855,0.725}
\definecolor{PeachPuff2}{rgb}{0.933,0.796,0.678}
\definecolor{PeachPuff3}{rgb}{0.804,0.686,0.584}
\definecolor{PeachPuff4}{rgb}{0.545,0.467,0.396}
\definecolor{NavajoWhite1}{rgb}{1.000,0.871,0.678}
\definecolor{NavajoWhite2}{rgb}{0.933,0.812,0.631}
\definecolor{NavajoWhite3}{rgb}{0.804,0.702,0.545}
\definecolor{NavajoWhite4}{rgb}{0.545,0.475,0.369}
\definecolor{LemonChiffon1}{rgb}{1.000,0.980,0.804}
\definecolor{LemonChiffon2}{rgb}{0.933,0.914,0.749}
\definecolor{LemonChiffon3}{rgb}{0.804,0.788,0.647}
\definecolor{LemonChiffon4}{rgb}{0.545,0.537,0.439}
\definecolor{cornsilk1}{rgb}{1.000,0.973,0.863}
\definecolor{cornsilk2}{rgb}{0.933,0.910,0.804}
\definecolor{cornsilk3}{rgb}{0.804,0.784,0.694}
\definecolor{cornsilk4}{rgb}{0.545,0.533,0.471}
\definecolor{ivory1}{rgb}{1.000,1.000,0.941}
\definecolor{ivory2}{rgb}{0.933,0.933,0.878}
\definecolor{ivory3}{rgb}{0.804,0.804,0.757}
\definecolor{ivory4}{rgb}{0.545,0.545,0.514}
\definecolor{honeydew1}{rgb}{0.941,1.000,0.941}
\definecolor{honeydew2}{rgb}{0.878,0.933,0.878}
\definecolor{honeydew3}{rgb}{0.757,0.804,0.757}
\definecolor{honeydew4}{rgb}{0.514,0.545,0.514}
\definecolor{LavenderBlush1}{rgb}{1.000,0.941,0.961}
\definecolor{LavenderBlush2}{rgb}{0.933,0.878,0.898}
\definecolor{LavenderBlush3}{rgb}{0.804,0.757,0.773}
\definecolor{LavenderBlush4}{rgb}{0.545,0.514,0.525}
\definecolor{MistyRose1}{rgb}{1.000,0.894,0.882}
\definecolor{MistyRose2}{rgb}{0.933,0.835,0.824}
\definecolor{MistyRose3}{rgb}{0.804,0.718,0.710}
\definecolor{MistyRose4}{rgb}{0.545,0.490,0.482}
\definecolor{azure1}{rgb}{0.941,1.000,1.000}
\definecolor{azure2}{rgb}{0.878,0.933,0.933}
\definecolor{azure3}{rgb}{0.757,0.804,0.804}
\definecolor{azure4}{rgb}{0.514,0.545,0.545}
\definecolor{SlateBlue1}{rgb}{0.514,0.435,1.000}
\definecolor{SlateBlue2}{rgb}{0.478,0.404,0.933}
\definecolor{SlateBlue3}{rgb}{0.412,0.349,0.804}
\definecolor{SlateBlue4}{rgb}{0.278,0.235,0.545}
\definecolor{RoyalBlue1}{rgb}{0.282,0.463,1.000}
\definecolor{RoyalBlue2}{rgb}{0.263,0.431,0.933}
\definecolor{RoyalBlue3}{rgb}{0.227,0.373,0.804}
\definecolor{RoyalBlue4}{rgb}{0.153,0.251,0.545}
\definecolor{blue1}{rgb}{0.000,0.000,1.000}
\definecolor{blue2}{rgb}{0.000,0.000,0.933}
\definecolor{blue3}{rgb}{0.000,0.000,0.804}
\definecolor{blue4}{rgb}{0.000,0.000,0.545}
\definecolor{DodgerBlue1}{rgb}{0.118,0.565,1.000}
\definecolor{DodgerBlue2}{rgb}{0.110,0.525,0.933}
\definecolor{DodgerBlue3}{rgb}{0.094,0.455,0.804}
\definecolor{DodgerBlue4}{rgb}{0.063,0.306,0.545}
\definecolor{SteelBlue1}{rgb}{0.388,0.722,1.000}
\definecolor{SteelBlue2}{rgb}{0.361,0.675,0.933}
\definecolor{SteelBlue3}{rgb}{0.310,0.580,0.804}
\definecolor{SteelBlue4}{rgb}{0.212,0.392,0.545}
\definecolor{DeepSkyBlue1}{rgb}{0.000,0.749,1.000}
\definecolor{DeepSkyBlue2}{rgb}{0.000,0.698,0.933}
\definecolor{DeepSkyBlue3}{rgb}{0.000,0.604,0.804}
\definecolor{DeepSkyBlue4}{rgb}{0.000,0.408,0.545}
\definecolor{SkyBlue1}{rgb}{0.529,0.808,1.000}
\definecolor{SkyBlue2}{rgb}{0.494,0.753,0.933}
\definecolor{SkyBlue3}{rgb}{0.424,0.651,0.804}
\definecolor{SkyBlue4}{rgb}{0.290,0.439,0.545}
\definecolor{LightSkyBlue1}{rgb}{0.690,0.886,1.000}
\definecolor{LightSkyBlue2}{rgb}{0.643,0.827,0.933}
\definecolor{LightSkyBlue3}{rgb}{0.553,0.714,0.804}
\definecolor{LightSkyBlue4}{rgb}{0.376,0.482,0.545}
\definecolor{SlateGray1}{rgb}{0.776,0.886,1.000}
\definecolor{SlateGray2}{rgb}{0.725,0.827,0.933}
\definecolor{SlateGray3}{rgb}{0.624,0.714,0.804}
\definecolor{SlateGray4}{rgb}{0.424,0.482,0.545}
\definecolor{LightSteelBlue1}{rgb}{0.792,0.882,1.000}
\definecolor{LightSteelBlue2}{rgb}{0.737,0.824,0.933}
\definecolor{LightSteelBlue3}{rgb}{0.635,0.710,0.804}
\definecolor{LightSteelBlue4}{rgb}{0.431,0.482,0.545}
\definecolor{LightBlue1}{rgb}{0.749,0.937,1.000}
\definecolor{LightBlue2}{rgb}{0.698,0.875,0.933}
\definecolor{LightBlue3}{rgb}{0.604,0.753,0.804}
\definecolor{LightBlue4}{rgb}{0.408,0.514,0.545}
\definecolor{LightCyan1}{rgb}{0.878,1.000,1.000}
\definecolor{LightCyan2}{rgb}{0.820,0.933,0.933}
\definecolor{LightCyan3}{rgb}{0.706,0.804,0.804}
\definecolor{LightCyan4}{rgb}{0.478,0.545,0.545}
\definecolor{PaleTurquoise1}{rgb}{0.733,1.000,1.000}
\definecolor{PaleTurquoise2}{rgb}{0.682,0.933,0.933}
\definecolor{PaleTurquoise3}{rgb}{0.588,0.804,0.804}
\definecolor{PaleTurquoise4}{rgb}{0.400,0.545,0.545}
\definecolor{CadetBlue1}{rgb}{0.596,0.961,1.000}
\definecolor{CadetBlue2}{rgb}{0.557,0.898,0.933}
\definecolor{CadetBlue3}{rgb}{0.478,0.773,0.804}
\definecolor{CadetBlue4}{rgb}{0.325,0.525,0.545}
\definecolor{turquoise1}{rgb}{0.000,0.961,1.000}
\definecolor{turquoise2}{rgb}{0.000,0.898,0.933}
\definecolor{turquoise3}{rgb}{0.000,0.773,0.804}
\definecolor{turquoise4}{rgb}{0.000,0.525,0.545}
\definecolor{cyan1}{rgb}{0.000,1.000,1.000}
\definecolor{cyan2}{rgb}{0.000,0.933,0.933}
\definecolor{cyan3}{rgb}{0.000,0.804,0.804}
\definecolor{cyan4}{rgb}{0.000,0.545,0.545}
\definecolor{DarkSlateGray1}{rgb}{0.592,1.000,1.000}
\definecolor{DarkSlateGray2}{rgb}{0.553,0.933,0.933}
\definecolor{DarkSlateGray3}{rgb}{0.475,0.804,0.804}
\definecolor{DarkSlateGray4}{rgb}{0.322,0.545,0.545}
\definecolor{aquamarine1}{rgb}{0.498,1.000,0.831}
\definecolor{aquamarine2}{rgb}{0.463,0.933,0.776}
\definecolor{aquamarine3}{rgb}{0.400,0.804,0.667}
\definecolor{aquamarine4}{rgb}{0.271,0.545,0.455}
\definecolor{DarkSeaGreen1}{rgb}{0.757,1.000,0.757}
\definecolor{DarkSeaGreen2}{rgb}{0.706,0.933,0.706}
\definecolor{DarkSeaGreen3}{rgb}{0.608,0.804,0.608}
\definecolor{DarkSeaGreen4}{rgb}{0.412,0.545,0.412}
\definecolor{SeaGreen1}{rgb}{0.329,1.000,0.624}
\definecolor{SeaGreen2}{rgb}{0.306,0.933,0.580}
\definecolor{SeaGreen3}{rgb}{0.263,0.804,0.502}
\definecolor{SeaGreen4}{rgb}{0.180,0.545,0.341}
\definecolor{PaleGreen1}{rgb}{0.604,1.000,0.604}
\definecolor{PaleGreen2}{rgb}{0.565,0.933,0.565}
\definecolor{PaleGreen3}{rgb}{0.486,0.804,0.486}
\definecolor{PaleGreen4}{rgb}{0.329,0.545,0.329}
\definecolor{SpringGreen1}{rgb}{0.000,1.000,0.498}
\definecolor{SpringGreen2}{rgb}{0.000,0.933,0.463}
\definecolor{SpringGreen3}{rgb}{0.000,0.804,0.400}
\definecolor{SpringGreen4}{rgb}{0.000,0.545,0.271}
\definecolor{green1}{rgb}{0.000,1.000,0.000}
\definecolor{green2}{rgb}{0.000,0.933,0.000}
\definecolor{green3}{rgb}{0.000,0.804,0.000}
\definecolor{green4}{rgb}{0.000,0.545,0.000}
\definecolor{chartreuse1}{rgb}{0.498,1.000,0.000}
\definecolor{chartreuse2}{rgb}{0.463,0.933,0.000}
\definecolor{chartreuse3}{rgb}{0.400,0.804,0.000}
\definecolor{chartreuse4}{rgb}{0.271,0.545,0.000}
\definecolor{OliveDrab1}{rgb}{0.753,1.000,0.243}
\definecolor{OliveDrab2}{rgb}{0.702,0.933,0.227}
\definecolor{OliveDrab3}{rgb}{0.604,0.804,0.196}
\definecolor{OliveDrab4}{rgb}{0.412,0.545,0.133}
\definecolor{DarkOliveGreen1}{rgb}{0.792,1.000,0.439}
\definecolor{DarkOliveGreen2}{rgb}{0.737,0.933,0.408}
\definecolor{DarkOliveGreen3}{rgb}{0.635,0.804,0.353}
\definecolor{DarkOliveGreen4}{rgb}{0.431,0.545,0.239}
\definecolor{khaki1}{rgb}{1.000,0.965,0.561}
\definecolor{khaki2}{rgb}{0.933,0.902,0.522}
\definecolor{khaki3}{rgb}{0.804,0.776,0.451}
\definecolor{khaki4}{rgb}{0.545,0.525,0.306}
\definecolor{LightGoldenrod1}{rgb}{1.000,0.925,0.545}
\definecolor{LightGoldenrod2}{rgb}{0.933,0.863,0.510}
\definecolor{LightGoldenrod3}{rgb}{0.804,0.745,0.439}
\definecolor{LightGoldenrod4}{rgb}{0.545,0.506,0.298}
\definecolor{LightYellow1}{rgb}{1.000,1.000,0.878}
\definecolor{LightYellow2}{rgb}{0.933,0.933,0.820}
\definecolor{LightYellow3}{rgb}{0.804,0.804,0.706}
\definecolor{LightYellow4}{rgb}{0.545,0.545,0.478}
\definecolor{yellow1}{rgb}{1.000,1.000,0.000}
\definecolor{yellow2}{rgb}{0.933,0.933,0.000}
\definecolor{yellow3}{rgb}{0.804,0.804,0.000}
\definecolor{yellow4}{rgb}{0.545,0.545,0.000}
\definecolor{gold1}{rgb}{1.000,0.843,0.000}
\definecolor{gold2}{rgb}{0.933,0.788,0.000}
\definecolor{gold3}{rgb}{0.804,0.678,0.000}
\definecolor{gold4}{rgb}{0.545,0.459,0.000}
\definecolor{goldenrod1}{rgb}{1.000,0.757,0.145}
\definecolor{goldenrod2}{rgb}{0.933,0.706,0.133}
\definecolor{goldenrod3}{rgb}{0.804,0.608,0.114}
\definecolor{goldenrod4}{rgb}{0.545,0.412,0.078}
\definecolor{DarkGoldenrod1}{rgb}{1.000,0.725,0.059}
\definecolor{DarkGoldenrod2}{rgb}{0.933,0.678,0.055}
\definecolor{DarkGoldenrod3}{rgb}{0.804,0.584,0.047}
\definecolor{DarkGoldenrod4}{rgb}{0.545,0.396,0.031}
\definecolor{RosyBrown1}{rgb}{1.000,0.757,0.757}
\definecolor{RosyBrown2}{rgb}{0.933,0.706,0.706}
\definecolor{RosyBrown3}{rgb}{0.804,0.608,0.608}
\definecolor{RosyBrown4}{rgb}{0.545,0.412,0.412}
\definecolor{IndianRed1}{rgb}{1.000,0.416,0.416}
\definecolor{IndianRed2}{rgb}{0.933,0.388,0.388}
\definecolor{IndianRed3}{rgb}{0.804,0.333,0.333}
\definecolor{IndianRed4}{rgb}{0.545,0.227,0.227}
\definecolor{sienna1}{rgb}{1.000,0.510,0.278}
\definecolor{sienna2}{rgb}{0.933,0.475,0.259}
\definecolor{sienna3}{rgb}{0.804,0.408,0.224}
\definecolor{sienna4}{rgb}{0.545,0.278,0.149}
\definecolor{burlywood1}{rgb}{1.000,0.827,0.608}
\definecolor{burlywood2}{rgb}{0.933,0.773,0.569}
\definecolor{burlywood3}{rgb}{0.804,0.667,0.490}
\definecolor{burlywood4}{rgb}{0.545,0.451,0.333}
\definecolor{wheat1}{rgb}{1.000,0.906,0.729}
\definecolor{wheat2}{rgb}{0.933,0.847,0.682}
\definecolor{wheat3}{rgb}{0.804,0.729,0.588}
\definecolor{wheat4}{rgb}{0.545,0.494,0.400}
\definecolor{tan1}{rgb}{1.000,0.647,0.310}
\definecolor{tan2}{rgb}{0.933,0.604,0.286}
\definecolor{tan3}{rgb}{0.804,0.522,0.247}
\definecolor{tan4}{rgb}{0.545,0.353,0.169}
\definecolor{chocolate1}{rgb}{1.000,0.498,0.141}
\definecolor{chocolate2}{rgb}{0.933,0.463,0.129}
\definecolor{chocolate3}{rgb}{0.804,0.400,0.114}
\definecolor{chocolate4}{rgb}{0.545,0.271,0.075}
\definecolor{firebrick1}{rgb}{1.000,0.188,0.188}
\definecolor{firebrick2}{rgb}{0.933,0.173,0.173}
\definecolor{firebrick3}{rgb}{0.804,0.149,0.149}
\definecolor{firebrick4}{rgb}{0.545,0.102,0.102}
\definecolor{brown1}{rgb}{1.000,0.251,0.251}
\definecolor{brown2}{rgb}{0.933,0.231,0.231}
\definecolor{brown3}{rgb}{0.804,0.200,0.200}
\definecolor{brown4}{rgb}{0.545,0.137,0.137}
\definecolor{salmon1}{rgb}{1.000,0.549,0.412}
\definecolor{salmon2}{rgb}{0.933,0.510,0.384}
\definecolor{salmon3}{rgb}{0.804,0.439,0.329}
\definecolor{salmon4}{rgb}{0.545,0.298,0.224}
\definecolor{LightSalmon1}{rgb}{1.000,0.627,0.478}
\definecolor{LightSalmon2}{rgb}{0.933,0.584,0.447}
\definecolor{LightSalmon3}{rgb}{0.804,0.506,0.384}
\definecolor{LightSalmon4}{rgb}{0.545,0.341,0.259}
\definecolor{orange1}{rgb}{1.000,0.647,0.000}
\definecolor{orange2}{rgb}{0.933,0.604,0.000}
\definecolor{orange3}{rgb}{0.804,0.522,0.000}
\definecolor{orange4}{rgb}{0.545,0.353,0.000}
\definecolor{DarkOrange1}{rgb}{1.000,0.498,0.000}
\definecolor{DarkOrange2}{rgb}{0.933,0.463,0.000}
\definecolor{DarkOrange3}{rgb}{0.804,0.400,0.000}
\definecolor{DarkOrange4}{rgb}{0.545,0.271,0.000}
\definecolor{coral1}{rgb}{1.000,0.447,0.337}
\definecolor{coral2}{rgb}{0.933,0.416,0.314}
\definecolor{coral3}{rgb}{0.804,0.357,0.271}
\definecolor{coral4}{rgb}{0.545,0.243,0.184}
\definecolor{tomato1}{rgb}{1.000,0.388,0.278}
\definecolor{tomato2}{rgb}{0.933,0.361,0.259}
\definecolor{tomato3}{rgb}{0.804,0.310,0.224}
\definecolor{tomato4}{rgb}{0.545,0.212,0.149}
\definecolor{OrangeRed1}{rgb}{1.000,0.271,0.000}
\definecolor{OrangeRed2}{rgb}{0.933,0.251,0.000}
\definecolor{OrangeRed3}{rgb}{0.804,0.216,0.000}
\definecolor{OrangeRed4}{rgb}{0.545,0.145,0.000}
\definecolor{red1}{rgb}{1.000,0.000,0.000}
\definecolor{red2}{rgb}{0.933,0.000,0.000}
\definecolor{red3}{rgb}{0.804,0.000,0.000}
\definecolor{red4}{rgb}{0.545,0.000,0.000}
\definecolor{DeepPink1}{rgb}{1.000,0.078,0.576}
\definecolor{DeepPink2}{rgb}{0.933,0.071,0.537}
\definecolor{DeepPink3}{rgb}{0.804,0.063,0.463}
\definecolor{DeepPink4}{rgb}{0.545,0.039,0.314}
\definecolor{HotPink1}{rgb}{1.000,0.431,0.706}
\definecolor{HotPink2}{rgb}{0.933,0.416,0.655}
\definecolor{HotPink3}{rgb}{0.804,0.376,0.565}
\definecolor{HotPink4}{rgb}{0.545,0.227,0.384}
\definecolor{pink1}{rgb}{1.000,0.710,0.773}
\definecolor{pink2}{rgb}{0.933,0.663,0.722}
\definecolor{pink3}{rgb}{0.804,0.569,0.620}
\definecolor{pink4}{rgb}{0.545,0.388,0.424}
\definecolor{LightPink1}{rgb}{1.000,0.682,0.725}
\definecolor{LightPink2}{rgb}{0.933,0.635,0.678}
\definecolor{LightPink3}{rgb}{0.804,0.549,0.584}
\definecolor{LightPink4}{rgb}{0.545,0.373,0.396}
\definecolor{PaleVioletRed1}{rgb}{1.000,0.510,0.671}
\definecolor{PaleVioletRed2}{rgb}{0.933,0.475,0.624}
\definecolor{PaleVioletRed3}{rgb}{0.804,0.408,0.537}
\definecolor{PaleVioletRed4}{rgb}{0.545,0.278,0.365}
\definecolor{maroon1}{rgb}{1.000,0.204,0.702}
\definecolor{maroon2}{rgb}{0.933,0.188,0.655}
\definecolor{maroon3}{rgb}{0.804,0.161,0.565}
\definecolor{maroon4}{rgb}{0.545,0.110,0.384}
\definecolor{VioletRed1}{rgb}{1.000,0.243,0.588}
\definecolor{VioletRed2}{rgb}{0.933,0.227,0.549}
\definecolor{VioletRed3}{rgb}{0.804,0.196,0.471}
\definecolor{VioletRed4}{rgb}{0.545,0.133,0.322}
\definecolor{magenta1}{rgb}{1.000,0.000,1.000}
\definecolor{magenta2}{rgb}{0.933,0.000,0.933}
\definecolor{magenta3}{rgb}{0.804,0.000,0.804}
\definecolor{magenta4}{rgb}{0.545,0.000,0.545}
\definecolor{orchid1}{rgb}{1.000,0.514,0.980}
\definecolor{orchid2}{rgb}{0.933,0.478,0.914}
\definecolor{orchid3}{rgb}{0.804,0.412,0.788}
\definecolor{orchid4}{rgb}{0.545,0.278,0.537}
\definecolor{plum1}{rgb}{1.000,0.733,1.000}
\definecolor{plum2}{rgb}{0.933,0.682,0.933}
\definecolor{plum3}{rgb}{0.804,0.588,0.804}
\definecolor{plum4}{rgb}{0.545,0.400,0.545}
\definecolor{MediumOrchid1}{rgb}{0.878,0.400,1.000}
\definecolor{MediumOrchid2}{rgb}{0.820,0.373,0.933}
\definecolor{MediumOrchid3}{rgb}{0.706,0.322,0.804}
\definecolor{MediumOrchid4}{rgb}{0.478,0.216,0.545}
\definecolor{DarkOrchid1}{rgb}{0.749,0.243,1.000}
\definecolor{DarkOrchid2}{rgb}{0.698,0.227,0.933}
\definecolor{DarkOrchid3}{rgb}{0.604,0.196,0.804}
\definecolor{DarkOrchid4}{rgb}{0.408,0.133,0.545}
\definecolor{purple1}{rgb}{0.608,0.188,1.000}
\definecolor{purple2}{rgb}{0.569,0.173,0.933}
\definecolor{purple3}{rgb}{0.490,0.149,0.804}
\definecolor{purple4}{rgb}{0.333,0.102,0.545}
\definecolor{MediumPurple1}{rgb}{0.671,0.510,1.000}
\definecolor{MediumPurple2}{rgb}{0.624,0.475,0.933}
\definecolor{MediumPurple3}{rgb}{0.537,0.408,0.804}
\definecolor{MediumPurple4}{rgb}{0.365,0.278,0.545}
\definecolor{thistle1}{rgb}{1.000,0.882,1.000}
\definecolor{thistle2}{rgb}{0.933,0.824,0.933}
\definecolor{thistle3}{rgb}{0.804,0.710,0.804}
\definecolor{thistle4}{rgb}{0.545,0.482,0.545}
\definecolor{gray5}{rgb}{0.051,0.051,0.051}
\definecolor{gray10}{rgb}{0.102,0.102,0.102}
\definecolor{gray15}{rgb}{0.149,0.149,0.149}
\definecolor{gray20}{rgb}{0.200,0.200,0.200}
\definecolor{gray25}{rgb}{0.251,0.251,0.251}
\definecolor{gray30}{rgb}{0.302,0.302,0.302}
\definecolor{gray35}{rgb}{0.349,0.349,0.349}
\definecolor{gray40}{rgb}{0.400,0.400,0.400}
\definecolor{gray45}{rgb}{0.451,0.451,0.451}
\definecolor{gray50}{rgb}{0.498,0.498,0.498}
\definecolor{gray55}{rgb}{0.549,0.549,0.549}
\definecolor{gray60}{rgb}{0.600,0.600,0.600}
\definecolor{gray65}{rgb}{0.651,0.651,0.651}
\definecolor{gray70}{rgb}{0.702,0.702,0.702}
\definecolor{gray75}{rgb}{0.749,0.749,0.749}
\definecolor{gray80}{rgb}{0.800,0.800,0.800}
\definecolor{gray85}{rgb}{0.851,0.851,0.851}
\definecolor{gray90}{rgb}{0.898,0.898,0.898}
\definecolor{gray95}{rgb}{0.949,0.949,0.949}
\definecolor{gray100}{rgb}{1.000,1.000,1.000}
\definecolor{DarkGray}{rgb}{0.663,0.663,0.663}
\definecolor{DarkBlue}{rgb}{0.000,0.000,0.545}
\definecolor{DarkCyan}{rgb}{0.000,0.545,0.545}
\definecolor{DarkMagenta}{rgb}{0.545,0.000,0.545}
\definecolor{DarkRed}{rgb}{0.545,0.000,0.000}
\definecolor{LightGreen}{rgb}{0.565,0.933,0.565}

\definecolor{ARed}{rgb}{1,0,0} 
\definecolor{BBlue}{rgb}{0.145,0.145,1} 
\definecolor{AllPurple}{rgb}{0.043,0.38,0.368} 

\newtheorem{Definition}{Definition}

\newcommand{\Maturity}{\textrm{Maturity}}
\newcommand{\Popularity}{\textrm{Popularity}}

\newcommand{\Size}{\textrm{Size}}

\begin{document}

\onehalfspacing

\title{Vulnerability Webs:\\ Systemic Risk in Software Networks\thanks{
Contact:  
\url{fritzc@tcd.ie}, \url{co.georg@fs.de}, \url{angelo.mele@jhu.edu}, and \url{mus47@psu.edu}. 
We wish to thank Bryan Graham, Stephane Bonhomme, Louise Laage, Matt Jackson, Eric Auerbach for helpful suggestions. 
We also thank participants at 2024 Cowles Foundation Conference in Econometrics, ESIF 2024 Conference on ML+AI, Empirical Methods in Game Theory Workshop, MEG 2024, EC24, MGOW 2024, Network Science in Economics 2023, 2023 BSE Summer Forum Networks Workshop, SEA 2023, Econometrics of Peer Effects and Networks, and seminar participants at Georgetown University, University of Washington, Virginia Tech, Boise State University, and the University of Pennsylvania for helpful comments and suggestions. 
The authors acknowledge support from the German Research Foundation award DFG FR 4768/1-1 (CF),
Ripple's University Blockchain Research Initiative (CPG),
the U.S.\ Department of Defense award ARO W911NF-21-1-0335 (MS, CF), 
and the U.S.\ National Science Foundation awards NSF DMS-1513644 and NSF DMS-1812119 (MS). 
Mele acknowledges support from an IDIES Seed Grant from the Institute for Data Intensive Engineering and Science at Johns Hopkins University and partial support from NSF grant SES 1951005.}
}
\date{}
\author[1]{Cornelius Fritz}
\affil[1]{Trinity College Dublin -- School of Computer Science and Statistics}
\author[2]{Co-Pierre Georg}
\affil[2]{{Frankfurt School of Finance and Management}}
\author[3]{Angelo Mele}
\affil[3]{{Johns Hopkins University -- Carey Business School}}
\author[4]{Michael Schweinberger}
\affil[4]{The Pennsylvania State University -- Department of Statistics}
\maketitle
\justify
\vspace{-2cm}

\begin{abstract}
Software development relies on code reuse to minimize costs, creating vulnerability risks through dependencies with substantial economic impact, as seen in the Crowdstrike and HeartBleed incidents. We analyze 52,897 dependencies across 16,102 Python repositories using a strategic network formation model incorporating observable and unobservable heterogeneity. Through variational approximation of conditional distributions, we demonstrate that dependency creation generates negative externalities. Vulnerability propagation, modeled as a contagion process, shows that popular protection heuristics are ineffective. AI-assisted coding, on the other hand, offers an effective alternative by enabling dependency replacement with in-house code. 


{\footnotesize {\bfseries Keywords:} economics of software development, dependency graphs, strategic network formation, exponential random graphs}

{\footnotesize {\bfseries JEL Classification:}  D85, L17, C31, L86, O31} 
\end{abstract}

\setcounter{page}{1}
\thispagestyle{empty}

\clearpage


%
\setcounter{page}{1}
%
\section{Introduction}\label{Sec::Introduction}
%
%



On July 19, 2024 a faulty update to the software Crowdstrike---a security software adopted by many companies---created  worldwide panic and disruptions, paralyzing domestic and international air traffic for hours, and disrupting operations at banks, hospitals, and hotel chains.\footnote{See \href{https://www.nytimes.com/2024/07/19/business/dealbook/tech-outage-crowdstrike-microsoft.html}{https://www.nytimes.com/2024/07/19/business/dealbook/tech-outage-crowdstrike-microsoft.html}} 
All told, the update affected 8.5 million Windows machines worldwide, 
each of which had to be manually rebooted to patch up the system and restart operations.
The faulty update of the Crowdstrike software was not the first instance highlighting systemic risk in software systems, and it will not be the last: e.g., in 2014 the Heartbleed bug---a vulnerability of the OpenSSL authentication library used by many websites---compromised the data of millions of users of websites such as Eventbrite, OkCupid, in addition to disrupting operations at the FBI and hospitals.
Such disruptions can generate staggering economic costs: Heartbleed, for example, resulted in at least $\$$500 million in damages.
According to the risk management company Parametrix, the disruption caused by Crowdstrike cost Fortune 500 companies $\$$5.4 billion.\footnote{See \cite{Parametrix2024}. }

The risk of these disruptions is deeply embedded in the structure of software systems. Modern software development is a collaborative effort that produces sophisticated software by making extensive use of already existing code \citep{Schueller2022}. This results in a complex network of dependencies among software packages, best described as dependency graphs.\footnote{Dependency graphs have been studied for many different programming languages using information from package managers. For an empirical comparison of the most common ones, see \cite{DecanMensGrosjean2019}.} While this re-use of code produces significant efficiency gains for software developers \citep{Lewis1991, Barros-Justo2018}, it also increases the risk that a vulnerability in one software package renders a large number of other packages equally vulnerable. In line with this view, the number of vulnerabilities listed in the Common Vulnerabilities and Exposure Database has increased almost twelve-fold between 2001 and 2019.\footnote{The Common Vulnerabilities and Exposure database is a public reference for known information-security vulnerabilities which feeds into the U.S.\ National Vulnerability Database. See \url{https://cve.mitre.org/}. The reported number of incidents has increased from 3,591 for the three years from 1999 to 2001 to 43,444 for the three-year period from 2017 to 2019.}
An estimate by \cite{CISQ2018} put the cost of losses from software failures at over \$1 trillion, 
up from \$59 billion in 2002 estimated by the \cite{NIST2002}.\footnote{Industry group Cybersecurity Ventures estimates that the damages incurred by all forms of cyber crime, including the cost of recovery and remediation, totalled USD 6 Trillion in 2021, and could reach USD 10.5 Trillion annually by 2025 (\href{https://cybersecurityventures.com/hackerpocalypse-cybercrime-report-2016/}{Source}).}  

Because vulnerabilities can potentially spread from one software package to all packages depending on it, coders can create an externality when deciding to re-use code from existing packages rather than implementing the required functionality themselves. 
Whether this is the case and how large the externality is is ultimately an empirical question.

We model coders' decisions to create a dependency to another software package as an equilibrium game of network formation with observed and unobserved heterogeneity.\footnote{Formally, we aggregate software packages that belong to the same code repository. Each package belongs to a single repository, but each repository can contain multiple packages. Repositories are typically managed by a small team of maintainers who decide which code is added to the repository.} 
Developers weigh the costs and benefits of forming links and revise these links at random times when there is a need for an update. 
Our equilibrium characterization provides the likelihood of observing a particular architecture of the software dependency network in the long run.
Using data on open source software projects contained in Python's software package manager Pypi, we obtain $52,897$ dependencies among $16,102$ repositories. 
We find evidence that a maintainer's decision to allow a dependency exerts a \emph{negative} externality on other maintainers. In principle, the presence of this externality discourages the formation of dependencies and partially contributes to the sparsity of the network.
Our result implies that it is particularly important to ensure that the code in these repositories is free from vulnerabilities that potentially affect the code in a large number of other repositories. 

We can use our model to study the effect of interventions that reduce code vulnerabilities that are likely to affect a large subset of the software ecosystem. 
We assume that vulnerabilities, coding errors, or bugs spread according to an epidemiological contagion model. Our simulations show that strategies to decrease the risk of contagion based on interventions targeting nodes according to measures of centrality or expected spread risk are not very effective in mitigating the average risk of contagion. 
On the other hand, we observe that the introduction of AI-assisted coding may decrease the incentives to form dependencies through a decrease in the costs of producing code. This process of decreasing the density of the network will mechanically decrease the average risk of contagion. We show that even moderate productivity gains from AI translate into a significant flattening of the risk profile for the software network.

Open-source software is an ideal laboratory to study software dependencies. Since the actual source code of software in a repository can be inspected and modified by others, code reuse is not only possible, it is encouraged. Modern software is not developed in isolation, but within an ecosystem of interdependent packages organized in code repositories. A package manager keeps track of code dependencies and ensures that these are automatically resolved so that when a user installs one package from a repository, all of the package's dependencies are also installed and automatically satisfied.
There are many programming languages that provide software using package managers. We focus on repositories of software packages written in the popular Python programming language and managed by the Pypi package manager.\footnote{ PyPI stands for Python Package Index  and is Python's official third-party software repository. See \href{https://pypi.org/help/}{here} for the official documentation.} Python is a modern programming language that is very popular in the software development community. 
We use data from \href{https://libraries.io}{libraries.io}, which, for each software repository, includes a full list of all dependencies.

We model the development of software as a process where maintainers reuse existing code, creating a network of dependencies among software packages---organized in repositories of several closely related software packages---in the process. This is typical for today's prevalent object-oriented software development paradigm. Our model describes a system of $N$ individual software repositories, characterized by observable and unobservable types. Each repository is managed by a single maintainer who decides which code to add to the repository and, consequently, which dependencies to form. 
Maintainers obtain a net benefit of linking to another package which depends on how active this dependency is maintained, how mature, popular, and large it is. We also allow a maintainers' utility to be affected by a local, i.e. type-specific, externality: Since dependent packages have dependencies themselves, they are susceptible to vulnerabilities imported from other packages. So we assume that maintainers care about the direct dependencies of the packages they link to. 
This specification excludes externalities that are more than two links away, as in the network formation models of \cite{DePaulaEtAl2014}, \cite{Mele2017} and \cite{MeleZhu2021}.
The network of dependencies forms over time and in each period a randomly selected package needs an update, so the maintainers decide whether to form a link or update the software in-house. Before updating the link, the package receives a random match quality shock. We characterize the equilibrium distribution over networks as a mixture of exponential random graphs \citep{SchweinbergerHandcock2015, Mele2022}, which can be decomposed into within- and between-types contribution to the likelihood. 

Model estimation is challenging because the likelihood depends on a normalizing constant that is infeasible to compute in large networks. Moreover, the model's unobserved heterogeneity has to be integrated out in the likelihood, making direct computations infeasible even for moderately sized networks. To alleviate these problems, we resort to a novel approximate two-step approach. In the first step, we estimate the discrete unobservable types of the nodes by approximating their conditional distribution via a fast variational mean-field approximation first proposed for stochastic blockmodels \citep{VuEtAl2013}. In the second step, we estimate the structural payoff parameters using a fast Maximum Pseudo-Likelihood Estimator (MPLE), conditioning on the estimated types \citep{BabkinEtAl2020, DahburaEtAl2021, DahburaEtAl2023}. We note that these methods were first developed for undirected graphs, and their application to directed networks necessitates non-trivial extensions of the computational machinery, providing an additional contribution of our work.\footnote{Our methods are implemented in the scalable, open-source, and platform-independent \texttt{R} package \texttt{bigergm}, which is available on CRAN servers \citep{cran}. Implementation details can be found in the appendix.} 

Our main result is that maintainers exert a \emph{negative} externality on other maintainers when creating a dependency on another repository: Other maintainers are then less likely to create a dependency on that repository. As a result,  the network is relatively sparse.
The interdependence among libraries implies that there exists a risk that a vulnerability in a single package has large adverse consequences for the entire ecosystem. An example of how a vulnerability in one package affected a significant part of critical web infrastructure is Heartbleed, the infamous bug in the widely used SSL/TLS cryptography library, resulting from incorrect validation of the input variable \citep{Durumeric2014}. When Heartbleed was disclosed in 2014, it made up to 55\% of secure webservers vulnerable to data theft, including the Canadian Revenue Agency and Community Health Systems, a large US hospital chain.
Our model provides a further argument for ensuring the security of highly interdependent software packages. Not only can a vulnerability in such a package affect a relatively large fraction of the entire ecosystem, 
because of the externality we identify,
it is also likely that this fraction increases as time goes on.

Another important question in the study of network formation processes is whether linked packages are similar or dissimilar with respect to their covariates. 
We find evidence of similarity among the same type of software packages in terms of their \Size, \Popularity \ and \Maturity  \ as well as between different-type software packages in terms of their \Popularity. The effect of similarity decreases with the (log of) the number of packages of the same unobservable type. 
We further find evidence of {\em dis-}similarity between different-type software packages in terms of their maturity and size.
In other words, mature, popular, and large software packages of one type are likely to depend on similar packages of the same type, but on less mature and smaller software packages of another type. 
This is intuitive because mature, popular, and large packages are likely to have a large user base with high expectations of the software, and coders try to satisfy this demand by providing sophisticated functionality with the help of mature, popular, and large software packages of the same type. These packages are often built using a lot of low-level functionality from large but fairly generic libraries, which is why large popular and mature software packages of one type depend on larger, but less popular and less mature software packages of another type.
One example of this is the inclusion of a payment gateway in an e-commerce application. The e-commerce application itself can be sophisticated and complex, like ebay is in the provision of their auction mechanism. For such an application, it is particularly valuable to provide users with the ability to use a variety of different payment methods, including credit card, paypal, or buy now pay later solutions like Klarna. 
Our result also aligns with the recent trend in software development away from large monolithic applications and towards interconnected microservices \citep{Traore2022}.

Lastly, we adapt a simple epidemiological contagion model to examine the spread of vulnerabilities in the Pypi dependency graph. We measure a repository's {\em $k$-step systemicness} as the number of downstream packages that are potentially rendered vulnerable by a vulnerability or bug in the upstream repository.
We use this model to study the efficacy of targeted interventions to mitigate vulnerability spread, drawing parallels with vaccination strategies in public health. By focusing on securing the most critical nodes--determined according to in-degree, expected fatality, or a combination of betweenness centrality and expected fatality--we assess the potential to significantly reduce the risk of vulnerability contagion. 

Our findings suggest that protecting even ten percent of the most critical repositories may reduce the average systemic risk but cannot completely flatten out the risk profile of the ecosystem.
On the other hand, we argue that the introduction of AI-assisted coding may have a greater effect on reducing systemic risk. Recent work has shown that AI tools such as GitHub Copilot can increase the productivity of software developers \citep{PengEtAl2023, EdelmanEtAl2023, NoyZhang2023}. In our model, this corresponds to a decrease in the cost of producing software in-house, without relying on creating dependencies or creating fewer dependencies. This can be modeled as an increase in the relative cost of a dependency. By simulating different magnitudes of the productivity improvement from AI-assisted coding, we show that the average systemic risk profile is flattened out by an increase of cost of merely five percent. The mechanism through which this risk reduction is achieved is the brute-force decrease in density for the network implied by increases in costs of dependencies. In general, our results help us to better understand the various driving forces and motifs in software development and how they shape the network of software dependencies. 

%
%
Our paper relates to several strands of literature in both economics and computer science.
First, our paper contributes to a growing literature on open source software, which has been an interest of economic research.\footnote{For an overview of the broad literature in the emerging field of digital economics, see \cite{GoldfarbTucker2019}.} In an early contribution, \cite{LernerTirole2002} argue that coders spend time developing open source software---for which they are typically not compensated---as a signal of their skills for future employers. Likewise, one reason why companies contribute to open source software is to be able to sell complementary services.
Open source projects can be large and complex, as \cite{Zheng2008} point out. They study dependencies in the Gentoo Linux distribution, and show that the resulting network is sparse, has a large clustering coefficient, and a fat tail. They argue that existing models of network growth do not capture the Gentoo dependency graph well and propose a preferential attachment model as alternative.\footnote{In earlier work, \cite{LaBelleWallingford2004} study the dependency graph of the Debian Linux distribution and show that it shares features with small-world and scale-free networks. However, \cite{LaBelleWallingford2004} do not strictly check how closely the dependency graph conforms with either network growth model.}

The package manager model is nowadays adopted by most programming languages which makes it feasible to use dependency graphs to study a wide variety of settings. \cite{Kikas2017}, for example, study the structure and evolution of the dependency graph of \verb|JavaScript|, \verb|Ruby|, and \verb|Rust|. The authors  emphasize that dependency graphs of all three programming languages become increasingly vulnerable to the removal of a single software package. An active literature studies the network structure of dependency graphs \citep{DecanMensGrosjean2019} to assess the vulnerability of a software ecosystem (see, for example, \citealp{Zimmermann2019}).

These papers show the breadth of literature studying open source ecosystems and dependency graphs. However, the literature considers the network either as stochastic or even as static and given. In contrast, we model the formation of dependencies as coders' {\em strategic} choice in the presence of various and competing mechanisms that either increase or reduce utility.\footnote{\cite{Blume2013} study how possibly contagious links affect network formation in a general setting. While we do not study the consequences of the externality we identify for contagion, this is a most worthwhile avenue for future research.}
Perhaps the closest work is \cite{Boysel2023}, that uses data from a sample of repositories to estimate a structural model of network formation and maintainers effort. In his model, maintainers sequentially update their dependencies, then play a network game where their efforts best responses are a function of other maintainers efforts. While related, our approaches are different, since \cite{Boysel2023} relies on sample panel data and the analysis focuses on the dynamics of myopic best-response. In our paper we focus on the long-run equilibrium and the cross-section of the dependencies among \emph{all} repositories at a particular point in time. 

The theoretical literature on strategic network formation has pointed out the role of externalities and heterogeneity in shaping the equilibrium networks \citep{bramoulle2016oxford, Galeotti2006, JacksonWolinsky1996}. 
Estimating strategic models of network formation is a challenging econometric task, because the presence of externalities implies strong correlations among links and multiple equilibria \citep{Mele2017, Snijders2002, DePaulaEtAl2014, Chandrasekhar2016, DePaula2017, BoucherMourifie2017, Graham2014, Graham2020}. In this paper, we model network formation as a sequential process and focus on the long-run stationary equilibrium  of the model \citep{Mele2017, Mele2022}. Because the sequential network formation works as an equilibrium selection mechanism, we are able to alleviate the problems arising from multiple equilibria.  

Adding unobserved heterogeneity further complicates identification, estimation and inference \citep{SchweinbergerHandcock2015, Graham2014, Mele2022}. Other works have considered conditionally independent links without externalities \citep{Graham2014, Graham2020, DePaula2017, Chandrasekhar2016}, providing a framework for estimation and identification in random and fixed effects approaches. On the other hand, because link externalities are an important feature in this context, we move away from conditionally independent links, and model nodes' unobserved heterogeneity as discrete types, whose realization is independent of observable characteristics and network, in a random effect approach. We can thus adapt methods of community discovery for random graphs to estimate the types, extending the work of \citet{BabkinEtAl2020} and in \citet{DahburaEtAl2021, DahburaEtAl2023} to accommodate for directed networks. 
Our two-steps method scales well to large networks, thus improving the computational challenges arising in estimation of these complex models \citep{BoucherMourifie2017, VuEtAl2013, BonhommeLamadonManresa2019}.

Our model is able to estimate the magnitude of externalities as well as homophily \citep{CurrariniJacksonPin2010, Jackson2008, Chandrasekhar2016}, the tendency of individuals to form links to similar individuals. Furthermore, our model is able to detect heterophily (or competition) among maintainers. More specifically, we allow homophily to vary by unobservables, while in most models homophily is estimated only for observable characteristics \citep{CurrariniJacksonPin2010, SchweinbergerHandcock2015, DePaulaEtAl2014, Chandrasekhar2016, Graham2020}.

Lastly, our contagion analysis is broadly related to the body of work that treats security as a strategic game on a given graph. The literature following \citet{GoyalVigier2014} and \citet{DziubinskiGoyal2013}, for example, shows how the optimal point of attack of an adversary and the defense strategy of a designer are based on the centrality of the network. Differently from this literature, though, we also study how changing the structure of the network, induced by a changing relative cost of forming dependencies, affects the spread of contagion and find this to be highly effective.





%
\section{A network description of code}\label{Sec::NetworkSoftware}
%

\subsection{The open source software paradigm}

The goal of computer programs is to implement algorithms on a computer. An algorithm is a terminating sequence of operations which takes an input and computes an output using memory to record interim results.\footnote{Memory to record interim instructions is sometimes called a ``scratch pad'', in line with early definitions of algorithms which pre-date computers. See, for example, Chapter 1 of \cite{Knuth1997-1}.} 
We use the term broadly to include algorithms that rely heavily on user inputs and are highly interactive (e.g. websites, spreadsheet and text processing software, servers).
Algorithms are implemented on a computer using code. Formally:
\begin{Definition} 
    Code is a sequence of operations 
    and arguments that implement an algorithm. A computer program is code that can be executed by a computer system. 
\end{Definition}
In order to execute a program, a computer provides resources---processing power and memory---and resorts to a compiler or interpreter, which in themselves are computer programs.\footnote{The term "compiler" was coined by \cite{Hopper1952} for her arithmetic language version 0 (A-0) system developed for the UNIVAC I computer. The A-0 system translated a program into machine code which are instructions that the computer can execute natively. Early compilers were usually written in machine code or assembly. Interpreters do not translate program code into machine code, but rather parse it and execute the instructions in the program code directly. Early interpreters were developed roughly at the same time as early compilers, but the first widespread interpreter was developed in 1958 by Steve Russell for the programming language LISP (see \cite{McCarthy1996}.}

Software developers, which we refer to as coders, use programming languages to implement algorithms. 
%
Today, the dominant modern software development paradigm is {\em object-oriented programming} (OOP).\footnote{For a principal discussion of object-oriented programming and some differences to procedural programming, see \cite{Abelson1996}. \cite{Kay1993} provides an excellent historical account of the development of early object-oriented programming.}  
Under this paradigm, code is developed primarily in {\em classes}, which contain data in the form of variables and data structures as well as code in the form of procedures.\footnote{Also called functions, or methods in the OOP paradigm.} Classes can communicate with one another via procedures using calls and returns. 
\begin{Definition}\label{Def::ClassObject} A class is a code template defining variables, procedures, and data structures. An object is an instance of a class that exists in the memory of a computer system.
\end{Definition}

Classes can interact in two ways. First, in the traditional {\em monolithic} software architecture, widely used for enterprise software like the one developed by SAP, for operating systems like Microsoft's Windows, and even in earlier versions of e-commerce platforms like Amazon, individual components cannot be executed independently. 
In contrast, many modern software projects use a {\em microservices} software architecture, which is a collection of cohesive, independent processes, interacting via messages.
Both architectures result in software where individual pieces of code depend on other pieces, either within a single application or across various microservices. These dependencies form a network which we formalize in the next section.

\subsection{Dependency Graphs}\label{Sec::NetworkSoftware:Dependency}
Most open source software is organized by package managers like PyPi for the Python programming language that standardize and automate software distribution. 
To make the installation of complex software projects easier for users, package managers keep track of a  package's dependencies and allow users to install these alongside the package they are installing. 
The existence of package managers and the level of automation they provide is necessary because modern software is frequently updated and different versions of the same package are not always compatible. 
Packages are logically grouped into repositories, controlled and managed by a maintainer.\footnote{Each package belongs to one repository, but a repository can contain multiple packages. Maintainers can also be organizations. Some repositories are also controlled by a group of maintainers, but for our purposes, this is immaterial.} 

Our unit of analysis is the network of dependencies among repositories. Specifically, we describe a software system consisting of $N$ repositories collected in $\mathcal{N} = \{1,\ldots,N\}$, each  managed by a different maintainer who decides whether to implement code themselves or to re-use existing code from other repositories. This (re-)use of existing code gives rise to linkages between repositories. 
The resulting network $\mathcal{G} \coloneqq (\mathcal{N},\, \mathcal{E})$ with $\mathcal{E} \subseteq \mathcal{N} \times \mathcal{N}$ is called the {\em dependency graph} of the software system.
We can represent the repository dependency graph $\mathcal{G}$ by a $N \times N$ adjacency matrix $g = \{ g_{ij} \}$, where $g_{ij}=1$ if $(i,j) \subseteq \mathcal{E}$ is an existing link from repo $i$ to repo $j$, indicating that $i$ depends on $j$. Since we consolidate code on the level of repositories, we exclude self-loops by defining $g_{ii} \coloneqq 0$.\footnote{This is not to say that there are no dependencies between code inside a repository--there will be many. But these are managed within a single repository by a maintainer. We are interested in the structure of the dependency network of repositories managed by different maintainers.}

There are two reasons for constructing the dependency graph on the repository rather than the package level. First, information about the popularity of a piece of software is relevant for maintainers linking decisions, but only created on the repository level.\footnote{For example, users can ``star'' a repository on GitHub if they find it particularly useful.} And second, repositories provide a logical grouping of closely related software packages. Alternatively, we could construct the dependency graph on the level of individual software packages. However, the breadth of code and functionality between packages can be much smaller than the breadth of code and functionality within a package.\footnote{Yet another alternative is to study {\em call graphs} arising from procedures within the same software package (see, for example, \citealp{Grove1997}). But these are state-dependent, i.e., dependencies arise during runtime and depending on how the package is executed and interacted with.}
%
Conveniently, the website \href{https://libraries.io/data}{libraries.io} provides information collected from various open source software package managers, including PyPi.\footnote{The website \href{https://libraries.io/data}{libraries.io} obtains this data by scraping publicly available repositories hosted on \href{https://github.com}{GitHub}, \href{https://gitlab.com}{GitLab}, and \href{https://bitbucket.org.}{Bitbucket}.}
The data provided includes information on projects--essentially the same as a PyPi package--as well as repositories and for each repository a list of all dependencies, defined on the project level. 
Since each project belongs to exactly one repository, we construct dependencies among repositories from the data provided.

On the repository level, this data includes information about the size of the repository in kB (Size), its popularity, measured as the number of stars it has on the website hosting the repository (Popularity), and as an alternative measure of popularity the number of contributors, i.e. the number of individual coders who have added code to a repository, raised or answered an issue, or reviewed code submitted by others.\footnote{We restrict ourselves to repositories that have a size above the 5\% percentile of the size distribution, and that have more than $2$ stars, but no more stars than the 95\% percentile of the distribution of stars.}

\begin{table}[t]
    \centering
    \begin{tabular}{cccccc}
    \toprule 
         \#Nodes & \#Edges & \#Components & \#Nodes (LCC) & \#Edges (LCC) \\
    \midrule 
        $17,095$ & $53,498$ & $422$ & $16,102$ & $52,897$ \\ 
    \bottomrule
    \end{tabular}
    \caption{Network statistics for the dependency graph of the PyPi ecosystem. }
    \flushleft \#Nodes and \#Edges is the number of nodes and edges for the total network and for the largest weakly connected component, respectively. \#Components is the number of connected components. 
    LCC indicates the largest (weakly) connected components.
    \label{Table::NetworkStats}
\end{table}

\begin{table}[t]
    \centering
    \begin{tabular}{llllllll}
    \toprule
            & Mean & Std.Dev. & Min. & p5 & Median & P95 & Max. \\
    \midrule
    In-Degree        & $3.29$    & $50.49$  & $0$     & $0$     & $0$   & $6$   & $4,210$ \\
    Out-Degree       & $3.29$    & $5.0$    & $0$     & $0$     & $2$   & $12$  & $95$    \\
    \bottomrule
    \end{tabular}
    \caption{In- and out-degree distribution.}\label{Tab::Centralities}
    \flushleft Distribution of in- and out-degree for the largest weakly connected component of the repo-based dependency graph of the PyPi ecosystem with $N=16,102$ nodes.
\end{table}

\begin{figure}[!ht]
    \includegraphics[width=0.9\textwidth]{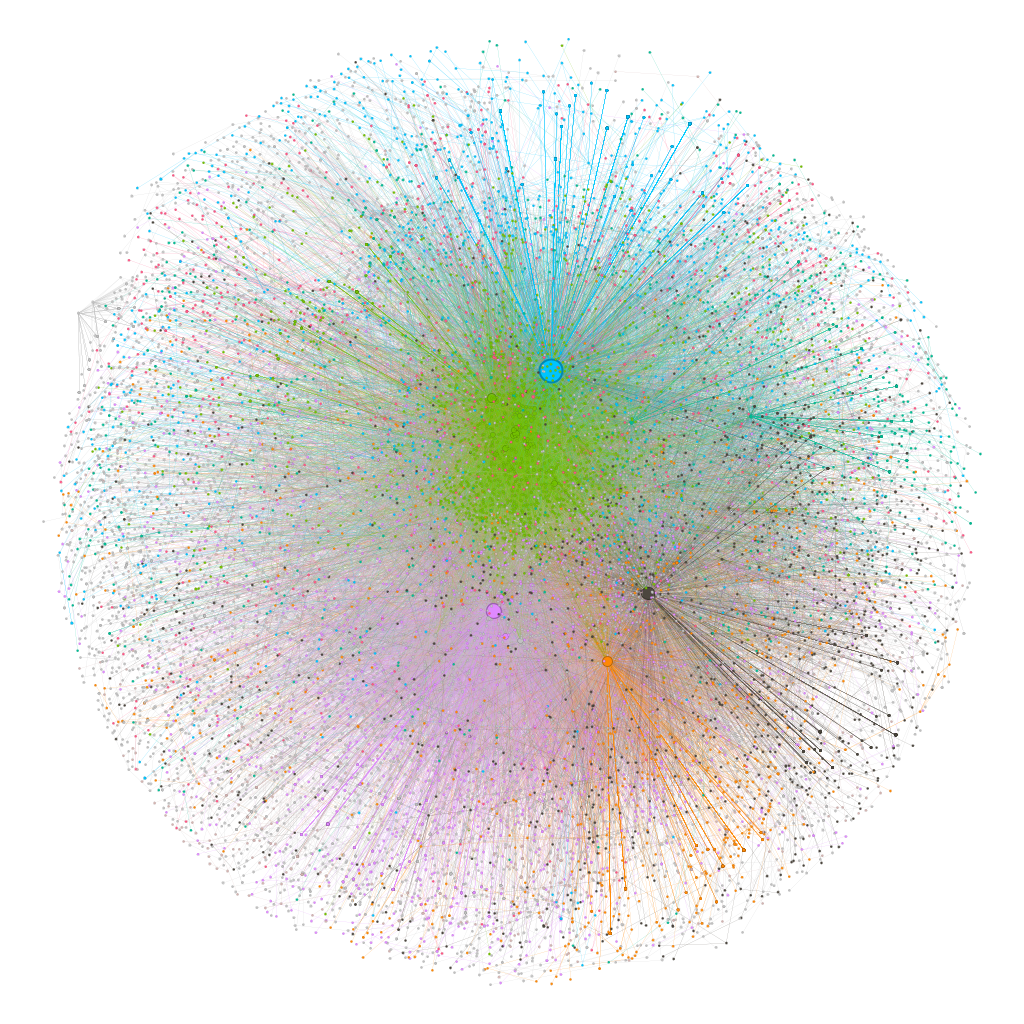}
    \caption{Dependency graph of the PyPi ecosystem.}
    \flushleft Nodes represent repositories in the \href{https://libraries.io}{libraries.io} dataset while edges represent dependencies of packages in different repositories. Colors indicate modularity classes according to \cite{Blondel2008}.
    \label{Fig::RustDependencyGraph}
\end{figure}

Table \ref{Table::NetworkStats} provides a high-level overview of the repository-based dependency graph constructed in this way. While there are $91$ weakly connected components, the largest weakly connected component covers almost all (over $99\%$) nodes and edges. This means that we still cover all relevant dynamics even by only focusing on the largest weakly connected component of the dependency graph. 

Furthermore, Table \ref{Tab::Centralities} shows that there is a lot of heterogeneity in how interconnected repositories are. The standard deviation of the out-degree, i.e. how many dependencies a given repository uses, is almost twice as large as the mean. Therefore, we observe substantial heterogeneity among repositories, with the median repository using $2$ dependencies and the maximum being $95$ dependencies. 
This heterogeneity is even more pronounced for the in-degree, i.e. how often a given repository is a dependency for another repository. 
The standard deviation is more than ten times as large as the mean; the median is $0$, while the maximum is $14,585$.

In Table \ref{Tab::TopRepos} we show the $10$ most depended-on repositories (i.e. with the highest in-degree). 
The repository with highest in-degree is \verb|six|, which facilitates compatibility between python version 2 and 3 (where some major updates have led to serious incompatibility issues). 
The second-highest in-degree repository is \verb|pytest-cov|, which is used to check a code's test coverage, which is a crucial step in modern software development. Likewise, the third-most commonly linked software repository is \verb|mock|, which is also used for software testing.

\begin{table}[tbp] \centering

\begin{tabular}{lccccc}

\toprule
\multicolumn{2}{c}{} & \multicolumn{3}{c}{Covariates}\\
\cmidrule{3-5} 
{Repo name} & In-Degree & {Popularity} & {Size} & {Maturity} \\
\midrule \addlinespace[\belowrulesep]
benjaminp/six                & 4,210 & 579 & 1,749 & 1,106 \tabularnewline
pytest-dev/pytest-cov        & 2,584 & 643 & 777   & 2,097 \tabularnewline
testing-cabal/mock           & 2,002 & 421 & 1,306 & 1,728 \tabularnewline
pypa/wheel                   & 1,572 & 156 & 1,302 & 902 \tabularnewline
kjd/idna                     & 1,336 & 106 & 507   & 2,422  \tabularnewline
pytest-dev/pluggy            & 952   & 364 & 471   & 1,721 \tabularnewline
pypa/packaging               & 782   & 141 & 540   & 2,069 \tabularnewline
pallets/markupsafe           & 775   & 282 & 175   & 3,492 \tabularnewline
PyCQA/mccabe                 & 767   & 242 & 82    & 2,518  \tabularnewline
coveralls-clients/coveralls-python & 761 & 406 & 328 & 2,532 \tabularnewline
\bottomrule
\end{tabular}
\caption{List of the ten most depended-on repositories for the PyPi ecosystem. }
\flushleft Popularity is measured as the number of stars a repository received on the website hosting it. Size is measured in Kilobytes of all code in the repository. Maturity  is the time in days between the first release of a repository and 13 January 2020.
\label{Tab::TopRepos} 
\end{table}

The largest weakly connected component of the repo-based dependency graph is shown in Figure \ref{Fig::RustDependencyGraph}, where we apply the community detection algorithm of \cite{Blondel2008} to color-code repositories in the same community. The three largest communities contain about $10-11\%$ of all repositories each. 


\subsection{Observables: Covariates}\label{Sec::NetworkSoftware:Covariates}
In addition to the software dependency data, we obtain additional data from the \url{libraries.io} website for each package and repository (e.g. \url{https://libraries.io/pypi/pandas}), which we show for the ten most depended-on repositories in Table \ref{Tab::TopRepos}.
We use three types of covariates, summarized in a vector of observable attributes $\bm{x}_i$.
First, we use a repository's \Size\ of all code in the repository, measured in Kilobytes. 
Second, 
\Popularity\ is measured as the number of stars--an expression of how many people like a repository or find it useful--a repository received on the website hosting it, 
e.g., 
\href{https://github.com}{GitHub}.
 And, third, the time in days between the first release of a repository and the reference date of 13 January 2020 is denoted $\Maturity$.

\begin{table}[t]
\begin{tabular}{lrrrrrrr}
\toprule 
& Mean & Std.dev & Min & p5 & Median & p95 & Max \\
\midrule
    Size [kB]       & 7,729.8  &  52,557.84    &      6   & 23 & 273 & 27,627 & 3,484,077 \\
    Popularity      & 70.49  &  133.35    &      3   &  3 & 16 & 364 & 853 \\
    Maturity & 1384.6 &  838.14 &   4  & 248  & 1250 & 2960  &  4270\\
\bottomrule 
\end{tabular}
\caption{Distribution of covariates.}
\flushleft The table reports some descriptives statistics of the covariates for the $N=16,102$ repositories in the largest connected component of our sample for the PyPi ecosystem.
Size is measured in Kilobytes of all code in the repository. 
Popularity is measured as the number of stars a repository received on the website hosting it.
Maturity  is the time in days between the first release of a repository and 13 January 2020.

\label{Tab::CovariatesDist}
\end{table}

To facilitate the estimation of our model, 
we create a categorical variable for each of our covariates using quartiles of the distribution. 
The reason for discretizing variables is computational, 
as our algorithm is based on stochastic block models with discrete types \citep{BickelEtAl2013, VuEtAl2013}. 
If the covariates are discrete, 
the estimation engine of stochastic block models can be adapted to estimate the unobserved heterogeneity \citep{VuEtAl2013, BabkinEtAl2020, DahburaEtAl2021}.
Otherwise,
when the covariates are continuous,
the computational costs are prohibitive for such large networks.
\section{Model}
\label{Sec::Model_dir}

We model software development as a dynamic process, assuming that maintainers update their repositories to add new functionality or improve existing functionality. Maintainers can update their repositories by adding functionality to their own code or creating dependencies to other repositories that provide the desired functionality.

\subsection{Setup}
\label{SubSec::Model_setup}

Each repository $i$ and its maintainer is characterized by a vector of $P$ observable attributes $\bm{x}_i \in \mathbb{R}^P$, 
which are stored in a matrix $\bm{x} \in \mathbb{R}^{N \times P}$. 
We assume that each of the $N$ repositories belongs to one of $K \geq 2$ types, 
unknown to researchers but known to other developers.
The type of a repository may be its purpose, 
which may be known to developers but may not be known to researchers,
or may correspond to unobservable qualities of the code and its developer team.
To indicate the type $k$ of repository $i$,
we define $z_{ik} \coloneqq 1$ and $z_{il} \coloneqq 0$ for all $l \neq k$ if repository $i$ is of type $k$,
and write $\bm{z}_i \coloneqq (z_{i1}, \ldots, z_{iK})$.
We assume that the unobservables are independent and identically distributed according to a multinomial distribution:
\begin{equation}
\label{multinomi}
\bm{Z}_i\; \overset{\mbox{\footnotesize iid}}{\sim}\; \mbox{Multinomial}\left(1;\, \bm\eta\right),\;\;
i = 1, \ldots, N,
\end{equation}
where $\bm\eta \coloneqq (\eta_1, \ldots, \eta_K)$ is a vector of $K$ probabilities summing to $1$.
We represent the \emph{software dependency network} $\bm{g}$ by the adjacency matrix, 
with elements $g_{ij} = 1$ if repository $i$ depends on repository $j$ and $g_{ij}=0$ otherwise. 
In general, 
software dependency networks are directed:
If repository $i$ depends on repository $j$,
then repository $j$ need not depend on repository $i$.

The utility that developers of repository $i$ receive from the software dependency network $\bm{g}$, 
with observables $\bm{x}$, 
unobservables $\bm{z}$,
and parameters $\bm\theta \coloneqq (\bm\alpha,\, \bm\beta,\, \gamma)$, 
is of the form
\begin{equation}
\label{EQ::Utility}
U_i(\bm{g};\, \bm{x},\, \bm{z},\, \bm\theta)\; 
\coloneqq\; \sum_{j=1}^N g_{ij}  u_{ij}(\bm\alpha,\, \bm\beta) + \sum_{j=1}^N\, \sum_{r\neq i,j}^N g_{ij} g_{jr}  v_{ijr}(\gamma) + \sum_{j=1}^N\, \sum_{r\neq i,j}^N g_{ij}  g_{ir}  w_{ijr}(\gamma).
\end{equation}

The function 
\begin{equation}
u_{ij}(\bm\alpha,\, \bm\beta)
\;\coloneqq\; u(\bm{x}_i,\, \bm{x}_j, \bm{z}_i,\, \bm{z}_j,\bm\alpha,\, \bm\beta)
\end{equation}
is the direct utility of linking repository $i$ to repository $j$. 
It is a function of observables ($\bm{x}_i,\, \bm{x}_j$), 
 unobservables ($\bm{z}_i,\, \bm{z}_j$),
 and parameters ($\bm\alpha,\, \bm\beta$).  
This represents the benefit of a dependency of repository $i$ on repository $j$, 
net of costs:
e.g.,
the developers will have to audit the code of repository $j$ and determine its quality,
and have to maintain the dependency on repository $j$.
The cost may include modification and adaptation of the code,
because developers must be able to use the functions and methods available in the repository $j$. 
Additional costs may arise from relying on other repositories,
because other repositories may contain undetected vulnerabilities.
Finally,
relying on existing code is convenient but may decrease the incentives of developers to maintain and improve coding skills,
which can undermine their future productivity and can hence be viewed as a cost.

Motivated by the application to software dependency networks,
we parameterize $u$ as follows:
\begin{equation}
u(\bm{x}_i, \bm{x}_j,\, \bm{z}_i, \bm{z}_j,\, \bm\alpha,\, \bm\beta)\; 
\coloneqq\; \alpha(\bm{z}_i, \bm{z}_j) + \displaystyle\sum_{p=1}^P \beta_{
p}(\bm{z}_i, \bm{z}_j)\, h_{ijp}(\bm{x}_i, \bm{x}_j), 
\end{equation}
where 
\begin{equation}
\begin{array}{lllll}
 \alpha(\bm{z}_i, \bm{z}_j) = \begin{cases}
      \alpha_{w1} + \alpha_{w1}^{\ell}\, \log(N_k), &\text{ if } \bm{z}_i = k \text{ and }  \bm{z}_j= k \\
      \alpha_b, &\text{ otherwise }\\
 \end{cases}
\end{array}
\end{equation}
and 
\begin{equation}
\begin{array}{lllll}
 \beta_{p}(\bm{z}_i, \bm{z}_j) = \begin{cases}
      \beta_{wp} + \beta_{wp}^{\ell}\, \log(N_k) ,&\text{ if } \bm{z}_i = k \text{ and }  \bm{z}_j= k \ \\
      \beta_{bp}, &\text{ otherwise }
 \end{cases}
\end{array}
\end{equation}
define within- and between-type utilities.

The terms in the within- and between-type utilities can be interpreted as follows: The intercept $\alpha_{w1} \in \mathbb{R}$ represents the fixed cost of forming a dependency;
the term $\alpha_{w1}^{\ell}\, \log(N_k)$ with weight $\alpha_{w1}^{\ell} \in \mathbb{R}$ is the part of the cost that depends on the size $N_k$ of type $k$;\footnote{In principle, the specification can be extended to incorporate aggregated terms at the type-level. 
In this sense, 
$\log(N_k)$ is one among other examples of terms that can be used to make the model more flexible. 
For instance, 
one may include the average of observed characteristics of type $k$ as one of the covariates. 
We do not pursue this specification here, given the very sparse nature of the software dependency network.}
and $h_{ijp}(\bm{x}_i, \bm{x}_j) \coloneqq \bm{1}\lbrace{x}_{ip} = {x}_{jp}\rbrace$ is an indicator function,
which is equal to $1$ if the $p$th covariate of $i$ and $j$ matches and is $0$ otherwise.
Size-dependent terms of the form $\log(N)$ were suggested by \citet{KrHaMo11}, 
motivated by invariance considerations (encouraging the expected number of links to be invariant to $N$),
and adapted to observed types by \citet{KrCoHe23} and to unobserved types by \citet{BabkinEtAl2020}.
The micro-behavioral foundations of size-dependent terms of the form $\log(N)$ were studied by \citet{butts:jms:2018}. The vectors $\bm\alpha \coloneqq (\alpha_{w1},\, \alpha_{w1}^{\ell},\, \alpha_b) \in \mathbb{R}^3$ and $\bm\beta \coloneqq (\beta_{w1},\, \ldots,\, \beta_{wP},\, \beta_{b1},\, \ldots,\, \beta_{bP}, \beta_{w1}^{\ell},\, \ldots,\, \beta_{wP}^{\ell}) \in \mathbb{R}^{3 P}$ are parameters to be estimated. 


The second and third term of utility function \eqref{EQ::Utility} are externalities generated by linking to repository $j$,
because repository $j$ may be linked to other repositories $r$ as well. 
This means that the developers of repository $i$ need to check the quality and features of the code in those repositories $r$ as well, 
to ensure that those repositories are free of vulnerabilities, 
and compatible with repository $i$.
On the other hand, 
any update of repository $r$ may compromise compatibility with repositories $i$ and $j$, 
increasing the cost of maintaining repository $i$.
In short, 
$w_{ijr}$ measures the net benefits of this externality when developers of repository $i$ form a dependency to repository $j$. 
In addition,
if the developers of repository $i$ create a dependency on repository $j$, 
it may compromise compatibility with other repositories $r$ that depend on repository $i$. 
This is accounted for by the third term $v_{ijr}$ in utility function \eqref{EQ::Utility}. 
We assume that the second and third term in the utility are of the form
\begin{equation}
   v_{ijr}(\gamma) 
   \;=\; w_{ijr}(\gamma)\;
   \coloneqq\; \begin{cases}
       \gamma  & \text{ if $i$, $j$, $r$ are all of type $k$}
       \\
       0 & \text{otherwise,}
   \end{cases}
\end{equation}
where $\gamma \in \mathbb{R}$.

The choice of the utility functions $U_i$ guarantees that all externalities are local, 
in the sense that all externalities are limited to repositories of the same type. 
Local externalities make sense in large networks,
because developers are unlikely to know the functionalities of thousands of other repositories and, 
more likely than not,
will create dependencies to other repositories of the same type,
which developers are most familiar with.
In addition,
local externalities have probabilistic, 
statistical,
and computational advantages.
First,
local externalities ensure desirable properties of models for large networks \citep{SchweinbergerHandcock2015}.
Second,
local externalities facilitate theoretical guarantees for statistical procedures,
which we discuss in Section \ref{sec4}.
Third,
local externalities enable local computing on subnetworks,
which facilitates parallel computing and hence large-scale computing \citep{BabkinEtAl2020}.

\subsection{Equilibrium}
\label{SubSec::Model_equil}

Software development is a dynamic process. 
We assume that time is discrete and that at each time point $t \in \{1, 2, \ldots\}$ a single repository is updated. In other words, maintainers do not update software repositories continuously, but certain events -- unobservable to researchers -- trigger the need for an update. For example, if repository $i$ depends on repository $j$, then an update of repository $j$ may require an update of repository $i$, or a request by users of repository $i$ may trigger an update of repository $i$. The update can be produced in-house or by forming a dependency to some other repository $j$ containing the same functionality. Since researchers may not be able to observe the timing of such updates, we model updates as a discrete-time Markov process \citep{Nj97}.

The first ingredient of the discrete-time Markov process is the update process that determines which repository $i$ requires an update and which dependency $j$ is considered to implement the update.
We assume that the probability that repository $i$ requires an update and the developers of repository $i$ consider creating, maintaining, or removing a dependence to repository $j$ can be expressed as a function $\rho_{i,j}$.
A simple example is an update process that selects an ordered pair of repositories $i$ and $j$ at random.
In other words,
a repository $i$ is selected with probability $1/n$ and requires an update.
Conditional on the event that repository $i$ requires an update, 
the developers of repository $i$ consider updating the possible dependency of repository $i$ on repository $j$ with probability $1/(n-1)$.
As a result,
the probability that repository $i$ requires an update and the developers of repository $i$ consider updating a (possible) dependency of repository $i$ on repository $j$ is
\[
\rho_{i,j}\left(\bm{g}^{t-1}, \bm{x}_i, \bm{x}_j, \bm{z}_i, \bm{z}_j\right)
\;\;=\;\; \dfrac{1}{n\, (n-1)},
\]
where $\bm{g}^{t-1}$ is the network at time $t-1$.
A second example is an update process that allows the conditional probability that an update of the (possible) dependency of repository $i$ on repository $j$ is proposed to depend on the distance $d(\bm{x}_i, \bm{x}_j)$ between the repositories $i$ and $j$.
The distance $d(\bm{x}_i, \bm{x}_j)$ is any distance function satisfying reflexity,
symmetry,
and the triangle inequality:
e.g.,
$d(\bm{x}_i, \bm{x}_j)$ may be the Euclidean distance between observable characteristics $\bm{x}_i$ and $\bm{x}_j$ of repositories $i$ and $j$,
that is,
$d(\bm{x}_i, \bm{x}_j) \coloneqq |\!|\bm{x}_i - \bm{x}_j|\!|_2$.
In other words,
a repository $i$ is selected with probability $1/n$ and,
conditional on the event that repository $i$ requires an update, 
the developers of repository $i$ consider updating the possible dependency of repository $i$ on repository $j$ with probability
$e^{-d(\bm{x}_i, \bm{x}_j)} / \sum_{k=1}^n e^{-d(\bm{x}_i, \bm{x}_k)}$,
which implies that
\[
\rho_{i,j}\left(\bm{g}^{t-1}, \bm{x}_i, \bm{x}_j, \bm{z}_i, \bm{z}_j\right)
\;\;=\;\; \dfrac{1}{n}\; \dfrac{e^{-d(\bm{x}_i, \bm{x}_j)}}{\sum_{k=1}^n e^{-d(\bm{x}_i, \bm{x}_k)}}.
\]

To characterize the stationary distribution of the discrete-time Markov process,
the update process needs to satisfy two mild conditions \citep{Mele2017}.
First,
the update probabilities $\rho_{i,j}$ 
can depend on the entire software dependency network excluding $(i, j)$:
\[
\rho_{i,j}\Big(\bm{g}^{t-1}, \bm{x}_i, \bm{x}_j, \bm{z}_i, \bm{z}_j\Big) 
\;\;=\;\; \rho_{i,j}\Big(\bm{g}_{-(i,j)}^{t-1}, \bm{x}_i, \bm{x}_j, \bm{z}_i, \bm{z}_j\Big).
\]
Second,
the update probabilities $\rho_{i,j}$ are strictly positive for all ordered pairs of repositories $(i, j)$:
\[
\rho_{i,j}\Big(\bm{g}_{-(i,j)}^{t-1}, \bm{x}_i, \bm{x}_j, \bm{z}_i, \bm{z}_j\Big)
\;\;>\;\; 0.
\]
This implies that each possible dependency can be proposed with a positive probability,
no matter how small.
These assumptions
help ensure that in the long-run equilibrium these probabilities do not affect the distribution of networks \citep{Mele2017}.

While the update process is stochastic, 
the decision to create,
maintain,
or remove a possible dependency of repository $i$ on repository $j$ is a strategic decision.
Conditional on the opportunity to update the dependency of repository $i$ on repository $j$,
the developers of repository $i$ make a decision about whether to create,
maintain,
or remove a dependency on repository $j$ based on utility function $U_i$. 
Before the developers of repository $i$ make a decision about the (possible) dependency on repository $j$,
the developers receive random shocks $\varepsilon_{ij0} \in \mathbb{R}$ and $\varepsilon_{ij1} \in \mathbb{R}$ to the perceived utility of a dependency of repository $i$ on repository $j$ ($g_{ij}^t=1$) and the perceived utility of writing code in-house ($g_{ij}^t=0$).
After receiving random shocks $\varepsilon_{ij0} \in \mathbb{R}$ and $\varepsilon_{ij1} \in \mathbb{R}$,
the developers of repository $i$ create a dependency on repository $j$ if the perceived utility of the dependency of repository $i$ on repository $j$ ($g_{ij}^t=1$) exceeds the utility of writing code in-house ($g_{ij}^t=0$):
\begin{equation}
 U_i\left(g_{ij}^{t}=1,\, \bm{g}_{-(i,j)}^{t-1};\, \bm{x}, \bm{z}, \bm{\theta}\right) + \varepsilon_{ij1}^t 
 \;\geq\; U_i\left(g_{ij}^{t}=0,\, \bm{g}_{-(i,j)}^{t-1};\, \bm{x}, \bm{z}, \bm{\theta}\right) + \varepsilon_{ij0}^t,   
\end{equation}
where $\bm{g}_{-(i,j)}^{t-1}$ refers to network $\bm{g}^{t-1}$ excluding the possible dependency $g_{ij}$ of repository $i$ on repository $j$. 
The random shocks to payoffs $\varepsilon_{ij0}^t \in \mathbb{R}$ and $\varepsilon_{ij1}^t \in \mathbb{R}$ capture the fact that unexpected events can occur when developing code in-house ($\varepsilon_{ij0}$) and when creating a dependency ($\varepsilon_{ij1}$), 
e.g., 
changes in the composition of the developer team or additional information about repository $j$. 
We assume that the random shocks to payoffs $\varepsilon_{ij0}$ and $\varepsilon_{ij1}$ do not depend on time,
are independent across pairs of repositories $(i, j)$,
and follow a logistic distribution, 
which is a standard assumption in the literature on discrete choice models \citep{Mele2017, GrahamDePaula2020, BoucherMourifie2017}.
As a consequence, 
when the opportunity to update the dependency of repository $i$ on repository $j$ arises,
the conditional choice probability of a dependency of repository $i$ on repository $j$ is
\begin{eqnarray}
    \mathbb{P}\left(g_{ij}^{t}=1 \ \vert \ \bm{g}_{-(i,j)}^{t-1};\, \bm{x}, \bm{z}, \bm{\theta}\right) = \frac{e^{U_i(g_{ij}^{t}=1,\, \bm{g}_{-(i,j)}^{t-1};\, \bm{x}, \bm{z}, \bm{\theta}) - U_i(g_{ij}^{t}=0,\, \bm{g}_{-(i,j)}^{t-1};\, \bm{x}, \bm{z}, \bm{\theta})}}{1 + e^{U_i(g_{ij}^{t}=1,\, \bm{g}_{-(i,j)}^{t-1};\, \bm{x}, \bm{z}, \bm{\theta}) - U_i(g_{ij}^{t}=0,\, \bm{g}_{-(i,j)}^{t-1};\, \bm{x}, \bm{z}, \bm{\theta})}}.
    \label{eq:CCP}
\end{eqnarray}

As shown in \citet[][]{Mele2022} (with unobservables) and \citet[][]{Bu09micro} and \citet{Mele2017} (without unobservables),
the sequence of networks generated by the process described above is a discrete-time Markov process \citep{Nj97},
conditional on the unobservable types $\bm{z}$.

The discrete-time Markov chain described above is an irreducible and aperiodic discrete-time Markov chain with finite state space and hence admits a unique limiting distribution $\pi$,
which is the stationary distribution of the Markov chain \citep{Nj97}.
The stationary distribution $\pi$ can be represented in closed form as an exponential-family distribution \citep{Mele2017},
that is,
an exponential-family random graph model \citep[][]{ergm.book} with local dependence \citep{SchweinbergerHandcock2015},
of the form
\begin{eqnarray}
\pi(\bm{g};\, \bm{x},\, \bm{z},\, \bm\theta)
&\coloneqq& \left(\prod_{k=1}^K \frac{e^{Q_{kk}(\bm{g}_{kk};\, \bm{x},\, \bm{z},\, \bm\theta)}}{c_{kk}(\bm{x},\, \bm{z};\, \bm\theta)}\right)\; \left(\prod_{l \neq k}^K\, \prod_{i=1}^{N_k}\, \prod_{j=1}^{N_l} \frac{e^{u_{ij}( \bm\alpha_b,\, \bm\beta_b)} }{1 + e^{u_{ij}(\bm\alpha_b,\, \bm\beta_b)}}\right).
\label{eq:StationaryDistribution}
\end{eqnarray}
The potential function $Q_{kk}$ of the within-type-$k$ distribution is defined as
\begin{equation}
    \begin{array}{lllll}
Q_{kk}\left(\bm{g}_{kk};\, \bm{x},\, \bm{z},\, \bm\theta\right)  
&\coloneqq& \displaystyle\sum_{i=1}^N\, \sum_{j\neq i}^N  u_{ij}\left(\bm{\alpha}_w, \bm{\beta}_w, \bm{x_i}, \bm{x}_j, \bm{z}_i, \bm{z}_j\right)  g_{ij}  z_{ik}  z_{jk}\vspace{.25cm}
\\
&+& \gamma\, \displaystyle\sum_{i=1}^N\, \sum_{j \neq i}^N\, \sum_{r \neq i,j}^N g_{ij}  g_{jr} z_{ik}  z_{jk} z_{rk},
\label{eq:potential_function}
\end{array}
\end{equation}
while the normalizing constant $c_{kk}$ of the within-type-$k$ distribution is
\begin{eqnarray}
c_{kk}(\bm{x},\, \bm{z},\, \bm{\theta}) 
&\coloneqq& \displaystyle\sum_{\bm{\omega}\, \in\, \mathcal{G}_{kk}} e^{Q_{kk}(\bm{\omega};\; \bm{x},\, \bm{z},\, \bm\theta)},
\label{eq:norm_constant_k}
\end{eqnarray}
where the sum is over all possible networks $g_{kk}^\prime \in \mathcal{G}_{kk}$ of type $k$.

The limiting and stationary distribution $\pi$ of the Markov chain represents the long-run distribution of the software development process, 
conditional on the types $\bm{z}$ of repositories. 
The first term in \eqref{eq:StationaryDistribution} represents the likelihood of dependencies among repositories of the same type,
whereas the second term represents the likelihood of dependencies between repositories of different types. 
The stationary distribution $\pi$ implies that the incentives of developers are captured by a potential function of the form
\begin{eqnarray}
    Q(\bm{g},\bm{x},\bm{z},\bm{\theta}) &=& \sum_{k=1}^K Q_{kk}\left(\bm{g}_{kk};\, \bm{x},\, \bm{z},\, \bm\theta\right)  + \sum_{k=1}^{K}\sum_{l \neq k}Q_{kl}\left(\bm{g}_{kl};\, \bm{x},\, \bm{z},\, \bm\theta\right),
\end{eqnarray}
where $Q_{kk}$ is defined in \eqref{eq:potential_function} while $Q_{kl}$ is defined by
\begin{eqnarray}
   Q_{kl}\left(\bm{g}_{kl};\, \bm{x},\, \bm{z},\, \bm\theta\right)\;\; \coloneqq\;\; \sum_{i=1}^N \sum_{j=1}^N g_{ij}\, u_{b}(\bm{x}_i, \bm{x}_j, \bm{z}_i, \bm{z}_j, \bm{\alpha_b}, \bm{\beta}_b)\, z_{ik}\, z_{jl}.
\end{eqnarray}

In this software development process,
each developer team optimally responds to the existing software dependency network. 
The Nash equilibria of the underlying potential game of the software development process are the networks that maximize the potential function $Q$ \citep{Mele2017,Mele2022, MondererShapley2006, Chandrasekhar2016, Blume2013}.
These are networks such that no developer team is willing to form an additional dependency or delete an existing dependency, 
when the need arises to update repository dependencies.
Upon inspecting the stationary distribution $\pi$,
it is evident that the Nash equilibria have the highest probability of being observed,
because the Nash equilibria maximize $ Q_{kl}$ and so maximize $\pi$.

Having said that,
Nash networks will be inefficient in the presence of externalities.
Consider the welfare function defined by the sum of the individual maintainers utilities
\begin{equation}
    \begin{array}{llllll}
    W(\bm{g}, \bm{x}, \bm{z}, \bm{\theta}) 
    &\coloneqq& \displaystyle\sum_{i=1}^N U_i(\bm{g}, \bm{x}, \bm{z}, \bm{\theta})
     &=& Q(\bm{g},\bm{x},\bm{z},\bm{\theta}) \,+\,\gamma\, \displaystyle\sum_{k=1}^K  \sum_{i=1}^N\, \sum_{j \neq i}^N\, \sum_{r \neq i,j}^N g_{ij}  g_{ri}  z_{ik}  z_{jk}  z_{rk}.
\end{array}
\end{equation}
The welfare function is not equal to the potential function $Q$ when $\gamma \neq 0$,
so the networks that maximize welfare may not maximize the potential function and are hence not Nash equilibria.

\section{Estimation and Empirical Results}
\label{sec4}

We infer the software development process by assuming that the observed software dependency network is a sample from the stationary distribution $\pi$ of the discrete-time Markov chain described in Section \ref{SubSec::Model_equil}.
We develop a scalable two-step algorithm by decoupling the estimation of the unobserved types from the structural parameter vector $\bm{\theta}$, which we describe Section \ref{sec:est}. 
Theoretical guarantees for type recovery and parameter recovery for each step are available:
e.g.,
theoretical guaranteees for type recovery can be found in \citet{ScKrBu17},
whereas theoretical guarantees for parameter recovery given types include consistency \citep{SchweinbergerStewart2020} and asymptotic normality \citep{St24} of maximum likelihood estimators,
and rates of convergence for maximum likelihood and pseudo-likelihood estimators \citep{StSc20}.
Empirical results are presented in Section \ref{secresults}.

\subsection{Two-Step Algorithm}
\label{sec:est}

We describe a scalable two-step algorithm for estimating directed network formation models with observable and unobservable types.
Note that scalable methods are available for undirected network formation models \citep{BabkinEtAl2020,DahburaEtAl2021}, 
but no scalable methods are available for directed network formation models.

The likelihood function of the parameter vectors $\bm\theta$ and $\bm\eta$ of the software development process based on a single observation of a software dependency network $\bm{g}$
is 
\begin{equation}
\label{eq:text1}
\begin{array}{lll}
\ell(\bm\theta,\, \bm\eta)
&\coloneqq& \displaystyle\log\sum_{\bm{z} \in \mathcal{Z}} p\left(\bm{z};\, \bm\eta\right)\, \pi(\bm{g};\, \bm{x},\, \bm{z},\, \bm\theta),
\end{array}
\end{equation}
where $\mathcal{Z}$ is the set of all $K^N$ possible configurations of types $\bm{z}$,\,
$p$ is the multinomial probability of $\bm{z}$ according to \eqref{multinomi},\,
and $\pi$ is the stationary distribution of $\bm{g}$ conditional on observables $\bm{x}$,
unobservables $\bm{z}$,
and the structural parameter vector $\bm\theta$.
Maximizing $\ell$ with respect to $\bm\theta$ and $\bm\eta$ is challenging,
because it involves a sum over all $K^N$ possible assignments of $N$ repositories to $K$ types,
which is exponential in $N$.
Worse, 
for each of the $K^N$ possible assignments of $N$ repositories to $K$ types, 
the stationary distribution $\pi$ is a function of an intractable normalizing constant,
which is the sum of between- and within-type sums,
each of which is exponential in the number of ordered pairs of repositories $N_k\, (N_k - 1)$ of type $k$ ($k = 1, \ldots, K$). 
Therefore, 
direct computation of the likelihood function is infeasible unless $N$ is small (e.g., $N \leq 10$).

To develop scalable approximations of likelihood-based inference,
we follow \citet{BabkinEtAl2020} and decouple the estimation of the unobserved types from the estimation of the structural parameter vector $\bm{\theta}$ using the following two-step algorithm: 
\begin{enumerate}
    \item[] \textbf{Step 1:} Estimate the types of repositories.
    \item[] \textbf{Step 2:} Estimate the structural parameter vector $\bm\theta$ conditional on the estimated types. 
\end{enumerate} 

\paragraph{Step 1}

We estimate the unobserved types of repositories as follows.

First,
we approximate the loglikelihood function $\ell$ by assuming that there are no externalities ($\gamma=0$),
which reduces the loglikelihood function of the directed network formation model to the loglikelihood function of a directed stochastic block model and allows us to tap into the large toolbox developed for stochastic block models.

Second,
we bound the loglikelihood function $\ell$ with $\gamma = 0$ from below by using Jensen's inequality.
Adapting the arguments of \citet{VuEtAl2013} and \citet{BabkinEtAl2020} from undirected to directed network formation models reveals that the resulting lower bound $\ell_{LB}$ of the likelihood function $\ell$ is
\begin{equation}
\label{eq:tractable_lb_text1}
\begin{array}{lllll}
\ell_{LB}(\bm{\theta}, \bm{\eta};\, \bm{\Xi}) 
&\coloneqq&\displaystyle\sum_{i=1}^{N} \sum_{j=1}^N \sum_{k=1}^{K}\sum_{l=1}^{K} \xi_{ik} \,\xi_{jl} \log \,\pi_{ij,kl}(g_{ij}, \bm{x}_{ij}) + \displaystyle\sum_{i=1}^{N}\sum_{k=1}^{K}\xi_{ik}\left( \log \eta_{k}-\log \xi_{ik}\right),
\end{array}
\end{equation}
where $\bm{\Xi} \coloneqq (\xi_{ik}) \in [0,1]^{N\times K}$ are auxiliary parameters,
$\eta_k$ is the prior probability of being of type $k$,\,
and $\pi_{ij,kl}$ is the conditional choice probability of a dependency of repository $i$ on repository $j$,
provided $i$ is of type $k$ and $j$ is of type $l$ with covariates $\bm{x}_{ij} = (\bm{x}_i, \bm{x}_j)$.
The value $\xi_{ik}$ serves as approximations of the posterior probability of repositories $i$ being of type $k$. 

Third,
we maximize the lower bound $\ell_{LB}$ with respect to $\bm{\Xi}$ to obtain the best bound on the loglikelihood function $\ell$.
Variational algorithms that directly maximize $\ell_{LB}$ with respect to $\bm{\Xi}$ risk being trapped in local maxima and can be time-consuming.
To facilitate maximization of $\ell_{LB}$,
we construct a surrogate function $M$,
which is more convenient to maximize than $\ell_{LB}$.
\citet{VuEtAl2013} and \citet{BabkinEtAl2020} propose the following surrogate function,
which can be obtained using the arithmetic-geometric mean inequality along with concavity properties of logarithms:
\begin{eqnarray}
\label{eq:minorizer_text}
M\left(\bm{\Xi};\, \bm{\theta}, \bm{\eta}^{(t)}, \bm{\Xi}^{(t)}\right) 
&\coloneqq & \sum_{i = 1}^{N}  \sum_{j \neq i}^{N} \sum_{k=1}^{K} \sum_{l=1}^{K} \left(\xi_{ik}^{2} \frac{\xi_{jl}^{(t)}}{2\xi_{ik}^{(t)}} + \xi_{jl}^{2} \frac{\xi_{ik}^{(t)}}{2\xi_{jl}^{(t)}}\right) \log \pi_{ij,kl}^{(t)}(g_{ij}, \bm{x}_{ij}) \\
&+& \sum_{i=1}^{N} \sum_{k=1}^{K} \xi_{ik} \left(\log\eta_{k}^{(t)} - \log\xi_{ik}^{(t)} - \frac{\xi_{ik}}{\xi_{ik}^{(t)}} + 1\right),\notag 
\end{eqnarray}
where $\xi_{ij}^{(t)}$ denotes the estimate of $\xi_{ij}$ at iteration $t$.
Maximizing \eqref{eq:minorizer_text} with respect to $\bm{\Xi}$ forces $\ell_{LB}$ uphill and can be implemented by quadratic programming using fast sparse matrix operations.
The resulting minorization-maximization (MM) algorithm is less prone to being trapped in local maxima than variational algorithm that directly maximize $\ell_{LB}$ with respect to $\bm{\Xi}$.

Finally,
we assign repository $i$ to type $\widehat{k} = \mbox{argmax}_k\; \widehat{\xi}_{i,k}$,
where $\widehat{\xi}_{i,k}$ is an approximation of the posterior probability of $i$ being of type $k$.

\paragraph{Step 2}

We estimate the structural parameter vector $\bm\theta$ by maximizing the pseudo-likelihood function of $\bm{\theta}$,
which is the product of conditional choice probabilities of links conditional on the types estimated in Step 1 \citep{BabkinEtAl2020, DahburaEtAl2023}. 
Maximum pseudo-likelihood estimators are more scalable than maximum likelihood estimators and are therefore preferred on computational grounds.

The resulting two-step algorithm is scalable,
in that it can handle small and large networks,
and it has been tested on undirected networks with a quarter of million units \citep{DahburaEtAl2021}.
Section \ref{sec4} provides theoretical guarantees for type and parameter recovery in Steps 1 and 2.

\subsection{Results}
\label{secresults}

\hide{
\begin{table}[!h]
\caption{Parameter estimates and standard errors.} 
\begin{center}
\begin{tabular}{lrr}
\hline
&Within&Between\tabularnewline
&Estimate&Estimate\tabularnewline
&(Std. Error)&(Std. Error)\tabularnewline\hline
Edges ($\alpha_1$)&$-5.801$&$-7.979$\tabularnewline
 &(0.276)&(0.006)\tabularnewline
$\log(n)\times$ Edges $\left(\alpha_1^\ell\right)$&$-0.533$&\tabularnewline
 &(0.031)&\tabularnewline
Externality ($\gamma$)&$-0.095$&\tabularnewline
 &(0.02)&\tabularnewline
Maturity ($\beta_1$)&1.791&$-0.222$\tabularnewline
 &(0.305)&(0.011)\tabularnewline
$\log(n) \times$ Maturity $\left(\beta_1^\ell\right)$&$-0.123$&\tabularnewline
 &(0.035)&\tabularnewline
Popularity ($\beta_2$)&1.944&0.118\tabularnewline
 &(0.305)&(0.01)\tabularnewline
$\log(n)\times$Popularity $\left(\beta_2^\ell\right)$&$-0.18$&\tabularnewline
 &(0.035)&\tabularnewline
Size ($\beta_3$)&1.114&$-0.07$\tabularnewline
 &(0.31)&(0.011)\tabularnewline
$\log(n)\times$Size$\left(\beta_3^\ell\right)$&$-0.122$&\tabularnewline
 &(0.036)&\tabularnewline
\hline
\end{tabular}\end{center}
\flushleft\scriptsize Estimates and standard errors are obtained by Maximum Pseudolikelihood (MPLE), conditioning on the estimated types in the first step. The number of unobservable types for the first step is $K=10$. Standard error for the estimates are in parenthesis.
\label{tbl:results_python}
\end{table}
}

\begin{table}[!tbp]
\caption{Parameter estimates and standard errors.\label{tbl:results_python}} 
\begin{center}
\begin{tabular}{lrr}
\toprule
&Within&Between\tabularnewline
&Estimate&Estimate\tabularnewline
&(Std. Error)&(Std. Error)\tabularnewline
Edges ($\alpha_1$)&$-$5.801&-7.979\tabularnewline
 &(0.276)&(0.006)\tabularnewline
$\log(n)\times$ Edges ($\alpha_2)$&$-$0.533&\tabularnewline
 &(0.031)&\tabularnewline
Externality ($\gamma$)&$-$0.095&\tabularnewline
 &(0.02)&\tabularnewline
Maturity ($\beta_1$)&1.791&-0.222\tabularnewline
 &(0.305)&(0.011)\tabularnewline
$\log(n) \times$ Maturity \big($\beta_1^\ell$\big)&$-$0.123&\tabularnewline
 &(0.035)&\tabularnewline
Popularity ($\beta_2$)&1.114&-0.07\tabularnewline
 &(0.31)&(0.011)\tabularnewline
$\log(n)\times$Popularity \big($\beta_2^\ell$\big)&$-$0.122&\tabularnewline
 &(0.036)&\tabularnewline
Size ($\beta_3$)&1.944&0.118\tabularnewline
 &(0.305)&(0.01)\tabularnewline
$\log(n)\times$Size \big($\beta_3^\ell$\big)&$-$0.18&\tabularnewline
 &(0.035)&\tabularnewline
\bottomrule
\end{tabular}\end{center}
\flushleft Estimates and standard errors are obtained by Maximum Pseudolikelihood (MPLE), conditioning on the estimated types in the first step. The number of unobservable types for the first step is $K=10$. Standard error for the estimates are in parenthesis.
\label{tbl:results_python}
\end{table}

We estimate a model with $K=10$ unobserved types and initialize the types allocation for our algorithm using the Walktrap algorithm.\footnote{As suggested by \citet{WainwrightJordan2008} and others,
we ran the algorithm with multiple initial guesses of the type allocations, 
but report the results that achieved the highest lower bound.}
While there is no established method to infer $K$ in this class of models, we have run the estimation with $K=5$ and $K=15$ for robustness, but in both cases the model's fit was worse than with $K=10$.

The results of our estimation are reported in Table \ref{tbl:results_python}. The first column shows estimates of the structural parameters for links of the same (unobservable) type, while the remaining column provides the estimates for links of different types. 

We estimate the externality ($\gamma$) to be negative, providing empirical evidence of the interdependence of links suggested by our strategic model. Essentially, our results provide evidence that when developers form connections, they consider the indirect impacts of their actions, at least partially. This behavior leads to a network architecture that might not be as efficient as one where a central planner aims to maximize the overall utility of all participants. Indeed, the externality encourages developers to establish fewer connections than would be ideal from an aggregate utility-maximization perspective. 
Maintainers see the number of dependencies of a repository they consider linking to as a cost, thus leading to linking decisions that decrease the density of the aggregate network. The negative externality has further implications for contagion, as we explain in the next section.

The parameters $\alpha_1$ and $\alpha_1^\ell$ govern the density of the network and are both estimated negative. 
The parameter $\alpha_1$ governing links between types is likewise estimated negative.
The magnitudes of the estimated parameters imply that---other things being equal---in equilibrium sub-networks of libraries of the same type tend to be less sparse than between-types. The sparsity increases with the number of libraries belonging to a particular type. One interpretation is that $\alpha$ measures costs of forming links, which is higher when there are more available libraries to choose from, possibly because of the need to monitor and evaluate more code when forming a decision. 

 By contrast, the parameters $\beta_1, \beta_2, \beta_3$ are all positive within types, indicating that developers are more likely to link to repositories of the same type of similar maturity, popularity and size. On the other hand, this homophily seems attenuated by the size of the type, as shown by the negative coefficients of the parameters $\beta_1^\ell, \beta_2^\ell, \beta_3^\ell$.

 When forming dependencies with repositories of a different type, the effect of the observables is negative for maturity and popularity, while positive for size. We interpret this as evidence of complementarities across types. 
 
One interpretation consistent with these results is that developers use popular features of same-type repositories to include in their own code but use infrastructure-style code from different-type repositories, which is not necessarily very popular. 

\subsection{Model Fit}

To assess model fit,
we follow the methodology developed in the literature on ERGM and HERGMs and simulate the model under the estimated parameters \citep{HuGoHa08,BabkinEtAl2020}. 
We compute network statistics on those simulated networks and compare them with the corresponding statistics evaluated on the observed network. 
If the model fits the data, most of the simulated network statistics should resemble their observed counterparts.

\begin{figure}
    \centering
    \includegraphics[width = 0.8\textwidth]{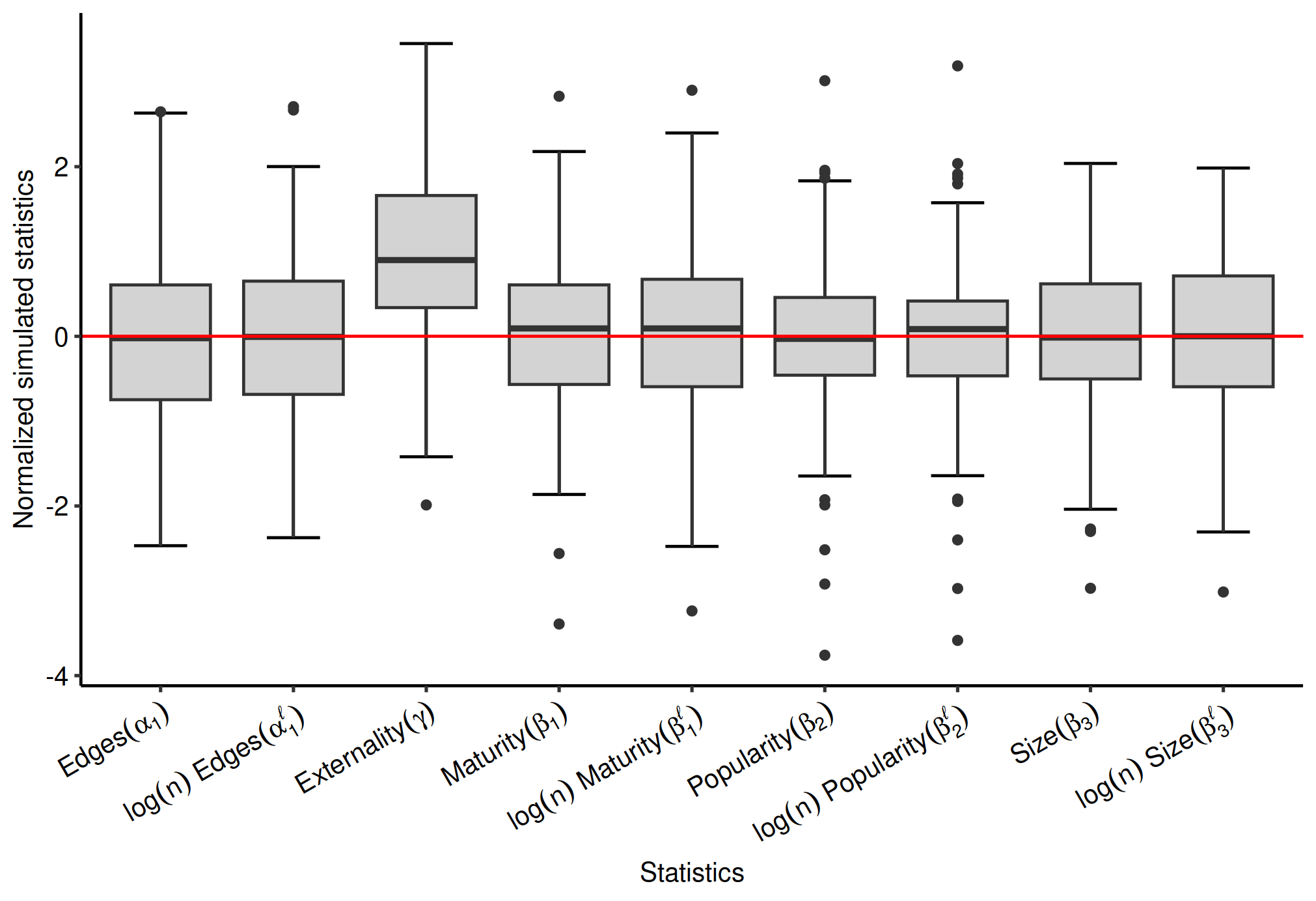}
    \caption{Model fit: Network statistics }
    \label{fig:gof_netstats_python}
\flushleft Each boxplot is based on 100 simulations of the network generated using the estimated parameters. Simulations are initialized at the observed network and proceed with a burn-in period of \mbox{500,000} Metropolis–Hastings steps. 
After burn-in, one network is retained every 50,000 steps. For each retained network, the corresponding network statistics are computed and then standardized such that a value of zero corresponds to the statistic of the observed network. 
On the x-axis, one boxplot is displayed for each network statistic associated with a coefficient of the fitted model.
\end{figure}

In Figure \ref{fig:gof_netstats_python} we show a boxplot with summaries of the networks obtained from 100 simulations of the model.
The summaries show in Figure \ref{fig:gof_netstats_python} are natural summaries, 
because they are the sufficient statistics of the model conditional on the types of repositories.
\footnote{We start the simulations at the observed network, and we use a Metropolis-Hastings sampler to update the network. We first run a burn-in of 500,000 iterations, and then sample a network every 50,000 iterations.}  
These summaries are normalized, 
so that a value of zero indicates a perfect match.
The red line is the observed network, so a perfect fit corresponds to all the network statistics concentrated around the red line. The boxplots show that the model is able to capture most of the aggregate structural properties of the Python dependency network: most of the simulated statistics are centered around the observed network. The only exception is the externality (i.e. the number of two-paths in our model), however the observed network falls in the $95\%$ confidence interval of the simulation output. We conclude that the estimated model does a good job in reproducing important features of the network.

\begin{figure}[ht]
    \centering
    \includegraphics[trim=0cm 0cm 0cm 0cm, clip, width=0.48\textwidth]{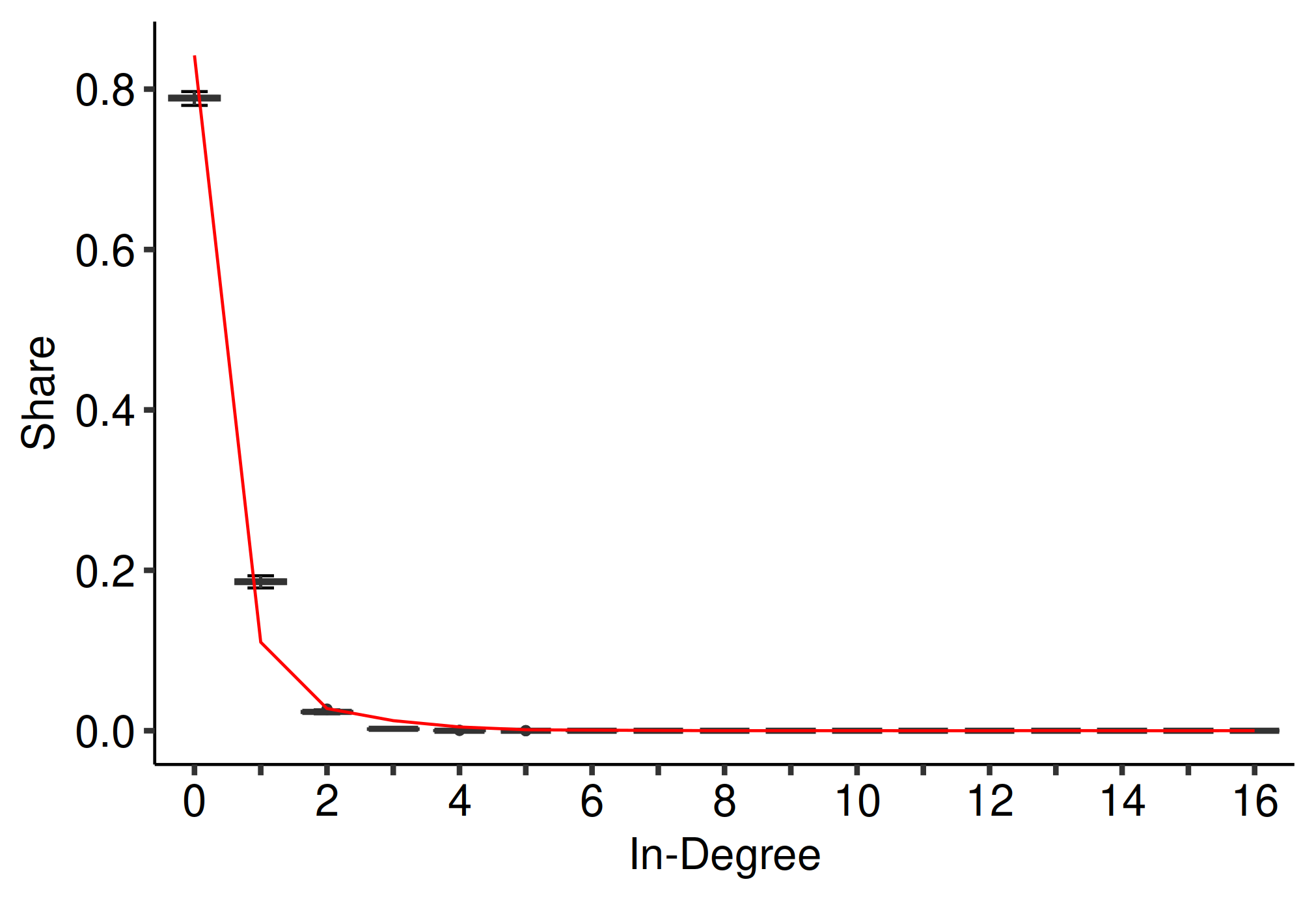}
    \includegraphics[trim=0cm 0cm 0cm 0cm, clip, width=0.48\textwidth]{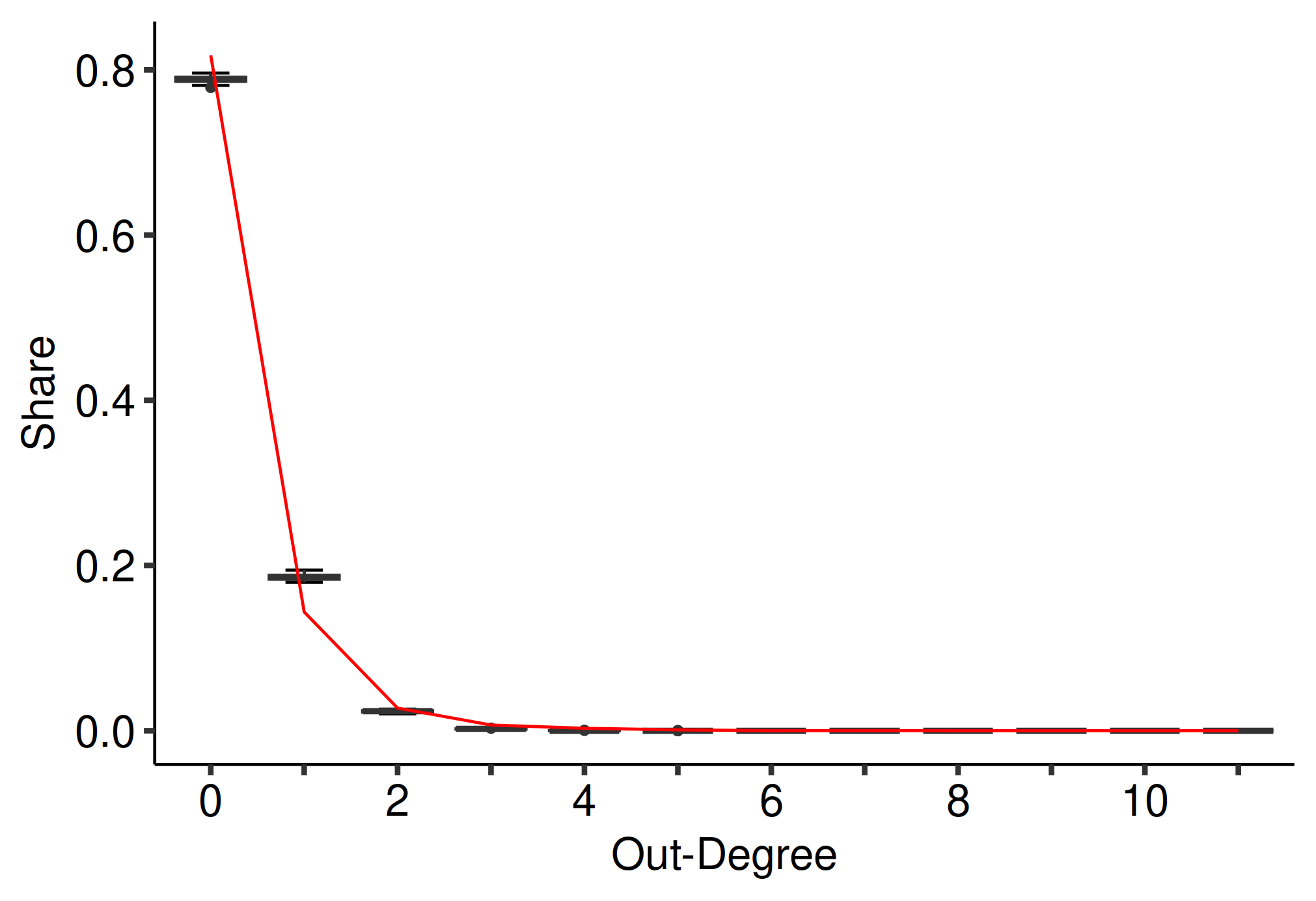}
 \caption{Model Fit: Degree distributions} 
    \label{fig:gof_degree_dist_python}
\flushleft The two plots represent the in-degree and out-degree distribution of the network. The red solid line is the observed degree distribution, while the (black) boxplots show the results of the simulations from our estimated model. 
\end{figure}

In Figure \ref{fig:gof_degree_dist_python} we report similar boxplots for in-degree and out-degree, showing that the simulated networks replicate the observed one quite closely. These results are particularly striking as it is usually quite hard to fit such a model for a very large network. 

In the Online Appendix \ref{Sec::Appendix:D}, we compare the fit of our model to alternative approaches. 
An important question is whether the unobserved heterogeneity included in our model is needed. To answer this question we compare our estimated model to a model without unobserved types ($K=1$). This corresponds to a standard Exponential Random Graph model (ERGM), as studied in \cite{Mele2017}. The ERGM is not able to match the aggregate density of the network, while our model's simulations are concentrated around the observed number of links. 
The second question is whether the externalities are important in explaining the topology of the equilibrium network. Therefore, we compare our model to a Stochastic blockmodel with covariates and $K=10$ types. This corresponds to our model with $\gamma = 0$. Our simulations show that the our model does a better job matching the externality within types, and should therefore be preferred. 
Finally, while there is no currently established method to choose the optimal number of types for this class of complex models, we have done a significant effort in setting an appropriate $K$ for estimation. Our practical approach consists of setting the number of types by checking the model's fit for  different values of $K$. In appendix \ref{Sec::Appendix:D} we show the fit with $K=5$ and $K=15$. 
These alternative models overestimate or underestimate the externality level, so our model with $K=10$ is preferred.

\section{Contagious Vulnerabilities}\label{Sec::Contagion}
%

A growing literature explores contagion on networks, including the statistical literature on infectious diseases (e.g., the contagion of HIV and coronaviruses). Such models \citep[see, e.g.,][]{ScBo20,GrWeHu11,BrNe02} usually first use a network formation model to generate a network of contacts among agents and then study how an infectious disease following a stochastic process
spreads within this network.

We adapt these ideas to the contagion of vulnerabilities in software dependency graphs. 
The prevalence of vulnerabilities in software motivated the creation of the Common Vulnerabilities and Exposure (CVE) program and the National Vulnerability Database (NVD) by the National Institute of Standards and Technology, which is an agency of the United States Department of Commerce \citep{NVDVulnerabilities}.
A software package depending on a vulnerable package is potentially vulnerable itself. To see how, consider, for example, the Equifax data breach \citep{FTC_EquifaxSettlement} in 2017, caused by a vulnerability in the popular open source web application framework Apache Struts 2.\footnote{Apache Struts 2 did not properly validate the Content-Type header of incoming http requests when processing file uploads, which allowed attackers to create http requests that include malicious code. Due to the lack of content validation, this allowed attackers to execute arbitrary code on the web server, ultimately granting them full control of the server. For details see the official NIST announcement on CVE-2017-5638 (\url{https://nvd.nist.gov/vuln/detail/CVE-2017-5638}).} 
Equifax did not itself develop or maintain Apache Struts 2, but instead used it as part of its own code base.
The vulnerability allowed attackers to execute malicious code providing access to the server hosting the web application developed by Equifax using Apache Struts 2 and costing the company more than \$1.5 Billion to date.\footnote{See \url{https://www.bankinfosecurity.com/equifaxs-data-breach-costs-hit-14-billion-a-12473}. Article accessed 27 January 2024.}

Package managers are part of a large suite of tools that help system administrators keep their systems up to date and patch any known vulnerabilities. It takes time for vulnerable systems to be updated \citep{Edgescan2022VulnerabilityStats},\footnote{It takes organizations between 146 and 292 days to patch cyber security vulnerabilities (Source: \href{https://www.statista.com/statistics/1363099/average-days-to-patch-vulnerability-by-severity/#:~:text=According%20to%20a%202022%20study,high%20severity%20within%20146%20days.}{Statista}. Accessed on 2024-01-31).} not all vulnerabilities are publicly disclosed \citep{Arora2008OptimalPolicy}, and it could even be optimal to release vulnerable software packages \citep{Arora2006SellFirst}. 
This leaves time for attackers to exploit vulnerabilities in downstream software packages and thereby attack the systems hosting code that depends on the vulnerable packages.

Since it is not ex ante clear which part of the dependency code is vulnerable, we assume that package $i$ uses vulnerable code in dependency $j$ with probability $\lambda_{ij}\in[0,1]$ and, for simplicity, we take $\lambda_{ij}=1$ and put ourselves in the worst case scenario. where an infected package would transmit the vulnerability with certainty to all downstream packages. This also avoids relying on statistical measures of the extent of contagion.

How contagious a vulnerability is can be measured in this setting by the number of downstream repositories it affects. While Github and other repository hosting services make it easy to see the number of packages directly depending on a given repository, direct dependencies are only imperfect proxies for the total number of vulnerable repositories two or three steps away. 
Formally, the d-neighborhood $\mathcal{N}^d_i(\bm{g})$ of node $i$ in network $\mathcal{G}$ is defined via:
\[\mathcal{N}_i^1(\bm{g}) = N_i(\bm{g}) \quad \textrm{and } \mathcal{N}^k_i(\bm{g}) = \mathcal{N}^{k-1}_i(\bm{g}) \cup \left( \bigcup_{j\in \mathcal{N}^{k-1}_i(\bm{g})} N_j(\bm{g}) \right), \]
where $g$ is the adjacency matrix of $\mathcal{G}$ and $N_i(\bm{g}) = \{j\in\mathcal{G}:\; g_{ji} = 1\}$ is the (in)-neighborhood of node $i$. 
Using this, we define a repository's {\em $k$-step systemicness} as $\textrm{Syst.}k_i(g) \equiv \mathcal{N}^k_i(g)$. We measure the aggregate {\em systemic risk} of the software dependency network under vulnerability contagion as the \emph{average} k-step systemicness of the network. 

Now that we have a measure for how ``infectious'' a vulnerable software package is, we turn to the question of how to prevent the spread of vulnerabilities.
We borrow from \cite{ChatterjeeZehmakan2023}, who compare different vaccination strategies in network-based contagion models. Since it is not feasible to find an optimal vaccination strategy--the optimization problem is NP hard--the authors compare various heuristics based on a node's network position and pathogen parameters. They find that a vaccination strategy based on a node's betweenness centrality and a heuristic they call {\em expected fatality} is most effective in preventing fatalities. 

By setting $\lambda_{ij}=1 \forall i,j\in N$, we assume that all upstream neighbors of a vulnerable node (repository) are also vulnerable. And second, nodes cannot ``die'' in our setup, so they are not removed from the population. They can, in principle, recover from being exposed to a vulnerable package if the vulnerability is patched and the dependency is updated, but this takes time, as discussed above. Consequently, software systems are vulnerable for a period of time once a vulnerability is discovered. We are interested in contagion processes occurring during this time. 

Consequently, we study the effectiveness of a prevention strategy based on securing the most important nodes against vulnerabilities (e.g. by through government funding or regulation). 
We compare three strategies. First, we rank the repositories based on their in-degree and target the nodes with highest indegree. Second, we follow \cite{ChatterjeeZehmakan2023} proposal to rank the nodes using an equally-weighted combination of a node's betweenness centrality and expected fatality $ef(i)$ of node $i$, which in our case is computed as:
\[ ef(i) = \sum_{j\in N(i)}\frac{1}{|N(j)|}.\]
Third, we use the expected fatality alone to rank the nodes. 

Using the ranking based on the indicators described above, we simulate our model and measure systemic risk. Our policy counterfactuals are obtained by simulating 1000 networks according to the model under the policy and then measuring the systemic risk according to the \emph{k-step systemicness}.\footnote{This procedure is essentially the same used for generating the model fit graphs, except that we change some parameters of the simulation to conform to the policies under study.} For each simulated network we record the mean of k-step systemicness as an aggregate measure of systemic risk. 

\subsection{Targeted Interventions}\label{Sec::Contagion:TargetedInterventions}

In the top left panel of Figure \ref{fig:policy} we show the effect of a policy that targets $1\% $ of the nodes and makes them \emph{invulnerable} to contagion. While this seems quite hard to achieve in practice, it is a good benchmark. We provide 3 different policies, targeting the nodes based on the highest in-degree, the expected fatality or a mix of betweenness centrality and expected fatality as explained above. The graph reports the model's average k-step systemicness in 1000 simulations (blue solid line). In equilibrium the average 5-steps systemicness for the Pypi ecosystem is close to 800 libraries. This implies that on average a bug or a vulnerability randomly introduced in the system will affect 800 libraries downstream in 5 steps. 

The policy targeting 1 percent of nodes based on the highest in-degrees is shown as purple dashed line; the policy based on the expected fatality is the green long-dash line; and the policy based on the indicator that equally weights betweenness centrality and expected fatality is the red dotted line.

\begin{figure}
    \caption{Mean k-step systemicness of policy targeting $1\%$, $5\%$ and $10\%$ of the nodes.}\label{fig:policy_all}
    \begin{minipage}{0.5\textwidth}
    \includegraphics[width=\textwidth]{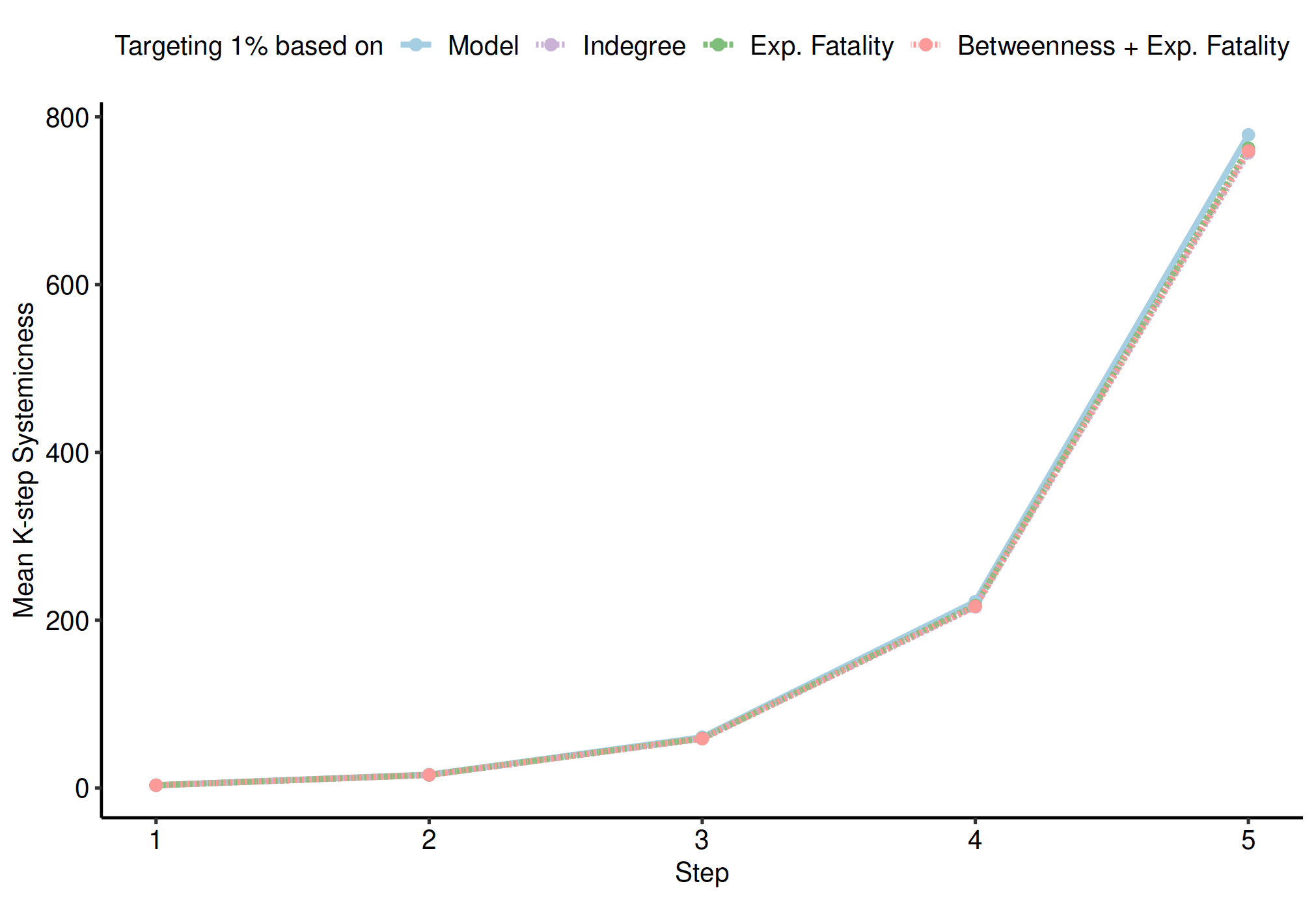}
    \end{minipage}\begin{minipage}{0.5\textwidth}
    \includegraphics[width=\textwidth]{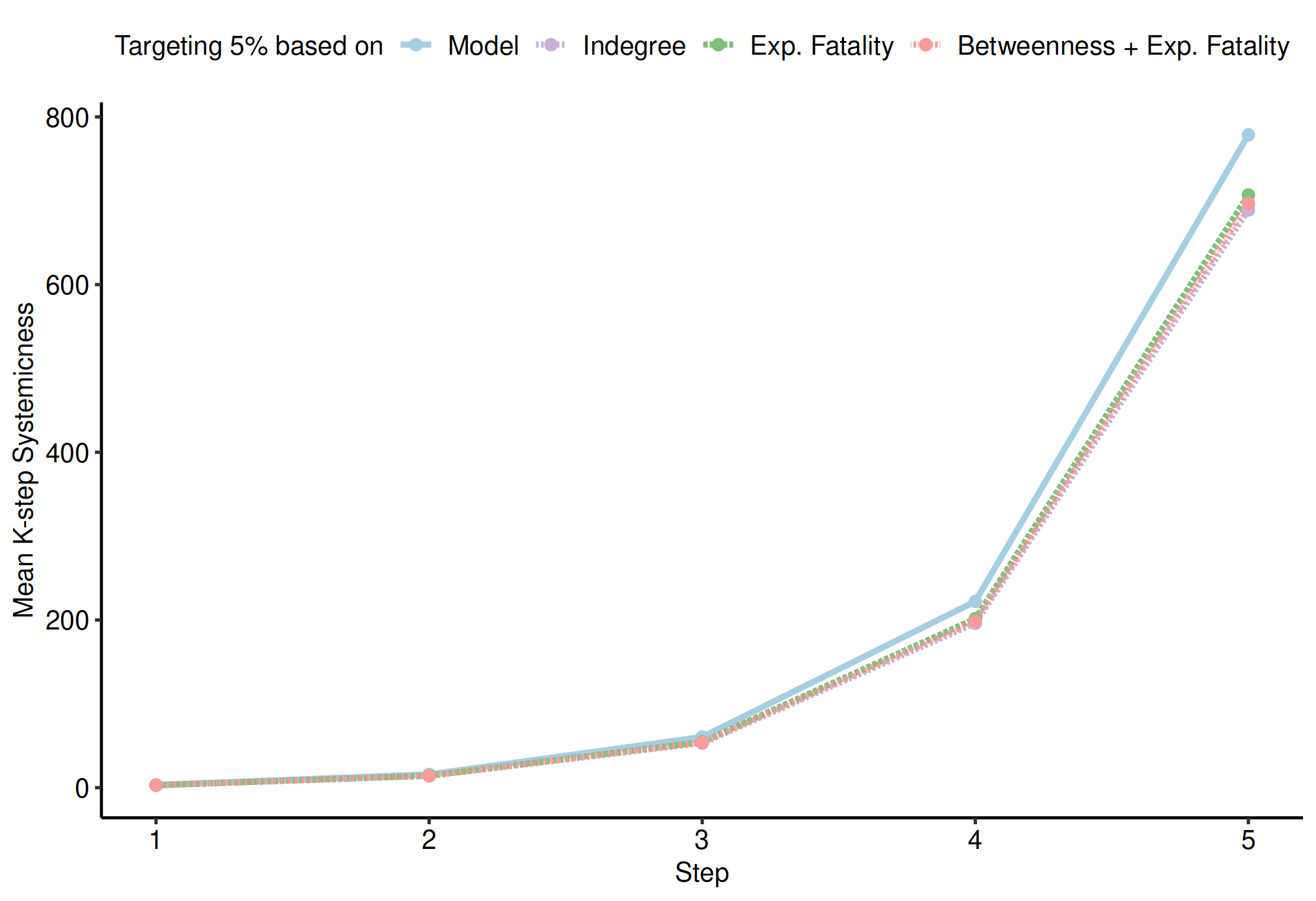}
    \end{minipage}
    \includegraphics[width=0.5\textwidth]{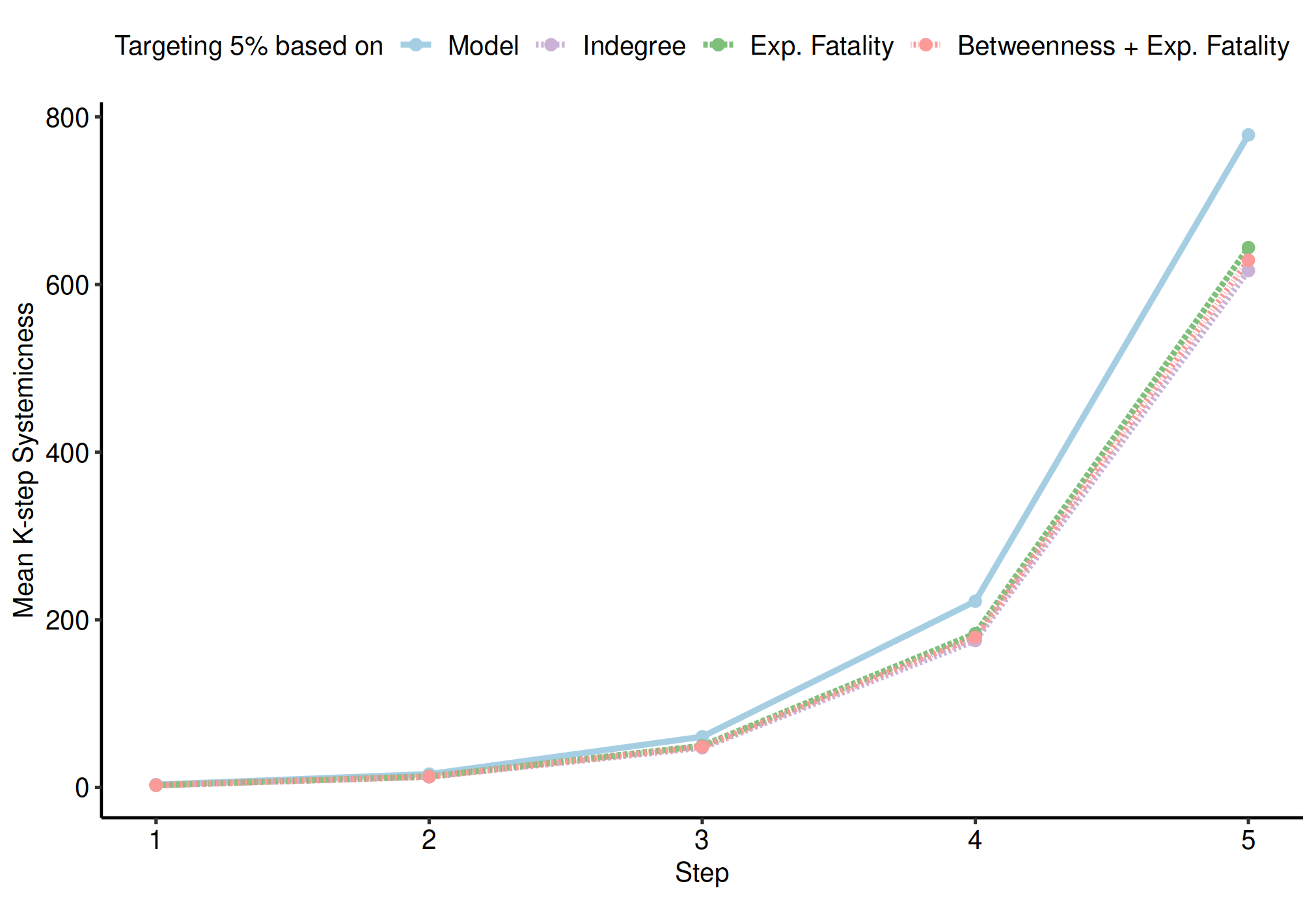}
    \flushleft Targeting is done according to In-degree, Expected Fatality, and $1/2\times$Expected Fatality +  $1/2\times$ Betweenness. Each line is based on the average of 1000 model simulations. Top left: Targeting based on top $1\%$ of nodes. Top right: Targeting based on top $5\%$ of nodes. Bottom: Targeting based on top $10\%$ of nodes. 
\end{figure}

This policy does not seem very effective. The systemic risk, as measured by the average \emph{k-step systemicness} of our simulations does not seem to change in a meaningful way.\footnote{We have performed t-tests to check whether the policy has significantly altered the average k-step systemicness and our test shows that this is the case, even if the effect is very small in this case.} We repeat the exercise in top right and bottom panel of Figure \ref{fig:policy_all}, where we target the highest $5$ and $10$ percent packages respectively, based on the targeting measure. While the effect of such policies is more pronounced, it does not really flatten out the risk profile of the network. 



\subsection{The effect of AI-assisted coding}\label{Sec::Contagion:AIAssist}
In recent years, the development of AI tools to help software development has increase exponentially. The introduction of LLM-based assistants has the potential to greatly increase productivity. In fact, recent studies about the adoption of such tools in software development support this claim. \cite{PengEtAl2023} provide experimental evidence that software developers using Github Co-Pilot were able to complete a task 55.8$\%$ faster, with less experienced developers benefiting the most from the AI-pair programming tool. \cite{EdelmanEtAl2023}, \cite{NoyZhang2023} and \cite{Yilmaz2023} find similar results in slightly different experimental settings.  

We contend that this development is likely to decrease systemic risk in the long-run. The effect of AI-assisted tools on software programming is as a first order decrease in cost of coding in-house. Each coder becomes more productive, and as a consequence the cost of producing high-quality software would decrease. In our model, this means that the relative cost of forming a dependency will increase, since our model payoffs measure the differential cost between producing in-house vs adapting the code to form a dependency. 
\begin{figure}
    \centering
    \includegraphics[width=0.8\linewidth]{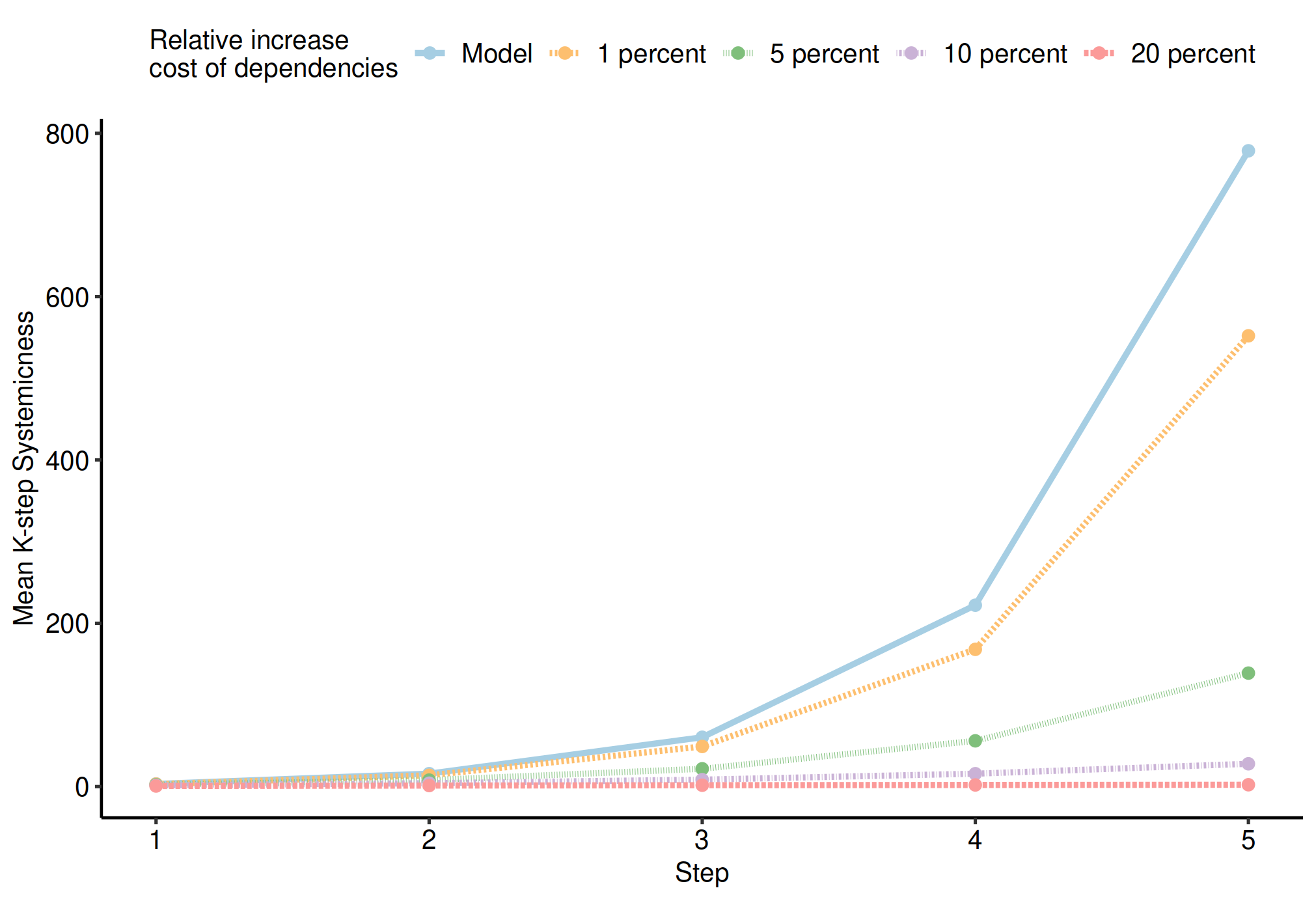}
    \caption{Mean k-step systemicness arising from introduction of AI-assisted coding.}
    \flushleft Each line is based on the average of 1000 model simulations of equilibrium adjustment, following a specific percent decrease in the cost of coding in-house.
    \label{fig:policy_c2}
\end{figure}

In light of this simple observation, we simulate the introduction of AI-assisted programming as an increase in cost of forming dependencies. We simulate different scenarios, where these costs increase by 1, 5, 10, 20 and 30 percent. 

These simulations are shown in Figure \ref{fig:policy_c2}. The model simulations are represented by the blue solid line. We notice that a small change in costs of 1 percent (yellow dashed line) has a pronounced effect on the systemic risk profile. A 5 percent change (green dotted line) dramatically improves the risk profile, thus being extremely effective. 
A 10 percent change (purple dotted dashed line) in the cost of forming a dependency essentially flattens out the systemic risk profile, as the network becomes relatively sparse and most coders do not form dependencies. 
This effect becomes stronger with a change of 20 percent in cost (represented as a red line).

\section{Conclusion\label{Sec::Conclusion}}

The network of dependencies among software packages is an interesting laboratory to study network formation and how incentives and externalities shape the topology of the network. Indeed, the creation of modern software gains efficiency by re-using existing libraries; on the other hand, dependencies expose new software packages to bugs and vulnerabilities from other libraries. This feature of the complex network of dependencies motivates the interest in understanding the incentives and equilibrium mechanisms driving the formation of such networks. 

In this paper, we estimate a directed network formation model to undertake a structural analysis of the motives, costs, benefits and externalities that a maintainer faces when developing a new software package. The empirical model allows us to disentangle observable from unobservable characteristics that affect the decisions to form dependencies to other libraries. 

Using data from the Python dependency network, we find evidence that coders create negative externalities for other coders when creating a link.
This raises more questions about the formation of dependency graphs. 
Using our estimated model, we study the vulnerability of the dependency network to spread of a bug. We measure systemic risk as the average number of downstream packages that are affected by a vulnerability, assuming that the contagion follows a SIR model of epidemic. We simulate interventions that target nodes according to their in-degree, expected fatality and betweenness centrality. Such interventions are unlikely to significantly decrease the risk profile of the network. On the other hand, the increase in AI-assisted tools for programmers, may have a beneficial effect by reducing the cost of producing code in-house, without much need for dependencies. 

While we have focused on a stationary realization of the network, there are important dynamic considerations in the creation of these dependencies. While the modeling of forward-looking maintainers may be useful to develop intuition about intertemporal strategic incentives and motives, we leave this development to future work, as it involves several complications in the estimation. Finally, we have focused on a single language, but the analysis can be extended to other languages as well, such as \verb|Java|, \verb|R|, \verb|Rust|, \verb|C++| and others.

%

\clearpage
\addcontentsline{toc}{section}{References}
\bibliographystyle{Meta/mod_agsm}
\bibliography{Meta/references}
\newpage

%
\appendix
\setcounter{table}{0}
\setcounter{figure}{0}
\renewcommand{\thetable}{A\arabic{table}}
\renewcommand{\thefigure}{A\arabic{figure}}
\clearpage

\newpage 
\begin{center}
\begin{LARGE}ONLINE APPENDIX\end{LARGE}
\end{center}
\section{Exploiting sparse matrix operations for updating variational parameters} 
\label{sec:expoiting}

Next, we detail how the updates of $\mathbf{\xi}$ in Step 1 can be carried out in a scalable manner based on sparse matrix operations. Equation \eqref{eq:tractable_lb_text1} can be written as a quadratic programming problem in $\mathbf{\xi}$ with the block constraints that $\sum_{k = 1}^K \xi_{ik} = 1$  needs to hold for all $i = 1, \ldots, N$. 
We rearrange the first line of \eqref{eq:tractable_lb_text1} as follows:
\begin{align*}
   & \sum_{i = 1}^{N} \sum_{j\neq i}^{N}  \sum_{k=1}^{K} \sum_{l=1}^{K} \left(\xi_{ik}^{2}\,\frac{\xi_{jl}^{(t)}}{2\xi_{ik}^{(t)}} + \xi_{jl}^{2} \, \frac{\xi_{ik}^{(t)}}{2\xi_{jl}^{(t)}}\right) \log \,\pi_{kl}^{(t)}(g_{ij}, \bm{x}_{ij})\\
    &= \sum_{i=1}^{N} \sum_{k=1}^{K} \sum_{j \neq i}^{N}  \sum_{l=1}^{K} \xi_{ik}^{2}\,\frac{\xi_{jl}^{(t)}}{2\xi_{ik}^{(t)}} \left( \log \,\pi_{kl}^{(t)} (g_{ij}, \bm{x}_{ij})+ \log \,\pi_{lk}^{(t)} (g_{ji}, \bm{x}_{ji})\right) \\
    &= \sum_{i=1}^{N} \sum_{k=1}^{K} \frac{\Omega_{ik}(\bm{g}, \bm{x}, \bm{\Xi}^{(t)})}{2\xi_{ik}^{(t)}}\,\xi_{ik}^{2}
\end{align*}
with 
\begin{align}
    \label{eq:omega}
    \Omega_{ik}\left(\bm{g}, \bm{x}, \bm{\Xi}^{(t)}\right) ~\coloneqq~ \sum_{j \neq i}^{N}  \sum_{l=1}^{K}  \xi_{jl}^{(t)} \left( \log \,\pi_{kl}^{(t)}(g_{ij}, \bm{x}_{ij})+ \log \,\pi_{lk}^{(t)} (g_{ji}, \bm{x}_{ji})\right).
\end{align}
We make the dependence of Equation \eqref{eq:omega} on $\bm{g}$ and $\bm{x}$ specific to ease the notation at later steps of this derivation.  
This yields for Equation \eqref{eq:tractable_lb_text1} 
\begin{align*}
M\left(\bm{\Xi}; \bm{\theta},\bm{\eta}^{(t)}, \bm{\Xi}^{(t)} \right) &= \sum_{i=1}^{N} \sum_{k=1}^{K}  \left(\frac{\Omega_{ik}^{(t)}\left(\bm{g}, \bm{x}, \bm{\Xi}^{(t)}\right)}{2\, \xi_{ik}^{(t)}} - \frac{1}{\xi_{ik}^{(t)}} \right)\xi_{ik}^{2} + 
\left(\log\eta_{k}^{(t)} - \log\xi_{ik}^{(t)} + 1\right)  \xi_{ik} \\ 
&= \sum_{i=1}^{N} \sum_{k=1}^{K} A_{ik}\left(\bm{g}, \bm{x},  \bm{\Xi}^{(t)} \right)\,\xi_{ik}^{2} + B_{ik}\left(\bm{\eta}^{(t)},  \bm{\Xi}^{(t)} \right)\,\xi_{ik}
\end{align*}
where 
\begin{align*}
    A_{ik}\left(\bm{g}, \bm{x},  \bm{\Xi}^{(t)}\right ) \coloneqq \frac{\Omega_{ik}\left(\bm{g}, \bm{x}, \bm{\Xi}^{(t)}\right)}{2\, \xi_{ik}^{(t)}} - \frac{1}{\xi_{ik}^{(t)}} 
\end{align*}
is the quadratic term and 
\begin{align*}
    B_{ik}\left(\bm{\eta}^{(t)}, \bm{\Xi}^{(t)} \right) \coloneqq \log\eta_{k}^{(t)} - \log\xi_{ik}^{(t)} + 1
\end{align*}
the linear term of the quadratic problem. 

To update the estimate of $\bm{\Xi}$, we need to evaluate $ A_{ik}\left(\bm{g}, \bm{x},  \bm{\Xi}^{(t)}\right )$ and $B_{ik}\left(\bm{\eta}^{(t)}, \bm{\Xi}^{(t)} \right)$ for $i = 1, \ldots, N$ and $k = 1, \ldots, K$, which, when done naively, is of complexity $O(N^2\, K^2)$. 
Computing $\Omega_{ik}\left(\bm{g}, \bm{x}, \bm{\Xi}^{(t)}\right)$ for $A_{ik}\left(\bm{g}, \bm{x},  \bm{\Xi}^{(t)}\right )$ is the computational bottleneck driving these complexities. 
It is thus prohibitively difficult for a large number of population members and communities in the network. Note that in our application, this problem is exasperated as we have more than 35,000 nodes and our theoretical result on the adequacy of the estimation procedure assumes that $K$ grows as a function of $N$. 
To avoid this issue, we show how $\Omega_{ik}\left(\bm{g}, \bm{x}, \bm{\Xi}^{(t)}\right)$ can be evaluated through a series of sparse matrix multiplications. 
Thereby, our implementation is fast and light on memory requirements.
The underlying idea is that $\Omega_{ik}\left(\bm{g}, \bm{x}, \bm{\Xi}^{(t)}\right)$ collapses to a simple form if $\mathbf{g}$ is an empty network with all pairwise covariates set to zero, i.e., $g_{ij} = 0$ and $x_{q} = 0\, \forall\, i,j = 1, \ldots, N$ and $q = 1, \ldots, p$. 
Given that, in reality, these are seldom the observed values, we, consecutively, go through all connections where either $g_{ij} = 1$ or $X_{ij,1} = x_{1}, \ldots, X_{ij,p} = x_{p}$ for $\sum_{q = 1}^p\, x_{q} \neq 0$ holds and correct for the resulting error. In formulae, this implies the following: 
\begin{align}
\label{eq:decomposition}
        \Omega_{ik}(\bm{g}, \bm{x}, \bm{\Xi}^{(t)}) ~&\coloneqq~ \sum_{j \neq i}^{N}  \sum_{l=1}^{K}  \xi_{jl}^{(t)} \left( \log \,\pi_{kl}^{(t)}(g_{ij}, \bm{x}_{ij})+ \log \,\pi_{lk}^{(t)} (g_{ji}, \bm{x}_{ji})\right) \notag\\ 
        &= \sum_{j \neq i}^{N}  \sum_{l=1}^{K}  \xi_{jl}^{(t)} \left( \log \,\pi_{kl}^{(t)} (0, \bm{0}) + \log \,\pi_{lk}^{(t)} (0, \bm{0})\right)\notag  \\ 
        &+\, 
        \sum_{j \neq i}^{N}  \sum_{l=1}^{K} \left(g_{ij} \xi_{jl}^{(t)} \left(\log \,\dfrac{\pi_{kl}^{(t)} (1, \bm{x}_{ij})}{\pi_{kl}^{(t)} (0, \bm{0})}\right) + g_{ji} \xi_{jl}^{(t)} \left(\log \,\dfrac{\pi_{lk}^{(t)} (1, \bm{x}_{ji})}{\pi_{lk}^{(t)} (0, \bm{0})}\right)\right) \notag\\ 
         &+\,
        \sum_{j \neq i}^{N}  \sum_{l=1}^{K} \left((1-g_{ij}) \xi_{jl}^{(t)} \left(\log \,\dfrac{\pi_{kl}^{(t)} (0, \bm{x}_{ij})}{\pi_{kl}^{(t)} (0, \bm{0})}\right) + (1-g_{ji}) \xi_{jl}^{(t)} \left(\log \,\dfrac{\pi_{lk}^{(t)} (0, \bm{x}_{ji})}{\pi_{lk}^{(t)} (0, \bm{0})}\right)\right) \notag\\ 
        &=  \Omega_{ik}\left(\bm{0}, \bm{0}, \bm{\Xi}^{(t)}\right) +  \Lambda_{ik}\left(\bm{g}, \bm{x},\bm{\Xi}^{(t)}\right), 
\end{align}
with 
\begin{align}
\label{eq:error}
   \Lambda_{ik}\left(\bm{g}, \bm{x}, \bm{\Xi}^{(t)}\right)&=         \sum_{j \neq i}^{N}  \sum_{l=1}^{K} \left(g_{ij} \xi_{jl}^{(t)} \left(\log \,\dfrac{\pi_{kl}^{(t)} (1, \bm{x}_{ij})}{\pi_{kl}^{(t)} (0, \bm{0})}\right) + g_{ji} \xi_{jl}^{(t)} \left(\log \,\dfrac{\pi_{lk}^{(t)} (1, \bm{x}_{ji})}{\pi_{lk}^{(t)} (0, \bm{0})}\right)\right)  \\
    &+ \sum_{j \neq i}^{N}  \sum_{l=1}^{K} \left((1-g_{ij}) \xi_{jl}^{(t)} \left(\log \,\dfrac{\pi_{kl}^{(t)} (0, \bm{x}_{ij})}{\pi_{kl}^{(t)} (0, \bm{0})}\right) + (1-g_{ji}) \xi_{jl}^{(t)} \left(\log \,\dfrac{\pi_{lk}^{(t)} (0, \bm{x}_{ji})}{\pi_{lk}^{(t)} (0, \bm{0})}\right)\right)\notag
\end{align}
being the error in $\Omega_{ik}\left(\bm{g}, \bm{x}, \bm{\Xi}^{(t)}\right)$ arising from assuming an empty network with all categorical covariates set to zero. 
 Next, we show how both terms, $ \Omega_{ik}\left(\bm{0}, \bm{0}, \bm{\Xi}^{(t)}\right) $ and $\Lambda_{ik}\left(\bm{g}, \bm{x},\bm{\Xi}^{(t)}\right)$, can be computed through matrix operations, completing our scalable algorithmic approach. 

We, first, assume an empty network where all categorical covariates set to zero, $g_{ij}= x_{q} = 0$  for all $i \neq j$ and $q= 1, ..., p$ and compute $\Omega_{ik}\left(\bm{0}, \bm{0}, \bm{\Xi}^{(t)}\right) $ via matrix multiplications. 
Observe that
\begin{align*}
     \Omega_{ik}\left(\bm{0}, \bm{0}, \bm{\Xi}^{(t)}\right)  &= \sum_{j \neq i}^{N}  \sum_{l=1}^{K}  \xi_{jl}^{(t)} \left(\log \,\pi_{kl}^{(t)}(0, \bm{0}) + \log \,\pi_{lk}^{(t)}(0, \bm{0})\right) \\
     &=  \sum_{l=1}^{K} \sum_{j \neq i}^{N} \xi_{jl}^{(t)} \left(\log \,\pi_{kl}^{(t)}(0, \bm{0}) + \log \,\pi_{lk}^{(t)}(0, \bm{0})\right) \\
     &= \sum_{l=1}^{K} \left( \sum_{j=1}^{N} \xi_{jl}^{(t)} - \xi_{il}^{(t)}\right) \left(\log \,\pi_{kl}^{(t)}(0, \bm{0}) + \log \,\pi_{lk}^{(t)}(0, \bm{0})\right) \\
     &= \sum_{l=1}^{K} \left(\tau^{(t)}_l - \xi_{il}^{(t)}\right)\left(\log \,\pi_{kl}^{(t)}(0, \bm{0}) + \log \,\pi_{lk}^{(t)}(0, \bm{0})\right)
\end{align*}
holds, where 
\begin{align*}
    \tau^{(t)}_l  ~\coloneqq~ \sum_{j=1}^{N} \xi_{jl}^{(t)}.
\end{align*}
With
\begin{align*}
    \bm{A_{0}}\left(\bm{\Xi}^{(t)}\right) ~\coloneqq~ \begin{bmatrix} 
   \tau^{(t)}_1 - \xi_{11}^{(t)} &\tau^{(t)}_2 - \xi_{12}^{(t)} & \dots & \tau^{(t)}_K - \xi_{1K}^{(t)} \\
   \tau^{(t)}_1 - \xi_{21}^{(t)} &\tau^{(t)}_2 - \xi_{22}^{(t)}  & \dots &\tau^{(t)}_K - \xi_{2K}^{(t)} \\
    \vdots & \vdots  & \ddots  & \vdots \\ 
   \tau^{(t)}(1) - \xi_{n1}^{(t)} & \tau^{(t)}_2 - \xi_{n2}^{(t)} & \dots  & \tau^{(t)}_K - \xi_{nK}^{(t)}
    \end{bmatrix}
\end{align*}
and
\begin{align*}
    \bm{\Pi_{0}}\left(\bm{\Xi}^{(t)}\right) ~\coloneqq~ \begin{bmatrix} 
    \log \,\pi_{11}^{(t)}(0, \bm{0}) & \log \,\pi_{12}^{(t)}(0, \bm{0}) & \dots & \log \,\pi_{1K}^{(t)}(0, \bm{0}) \\
    \log \,\pi_{21}^{(t)}(0,\bm{0}) & \log \,\pi_{22}^{(t)}(0,\bm{0}) & \dots & \log \,\pi_{2K}^{(t)}(0, \bm{0}) \\
    \vdots & \vdots  & \ddots  & \vdots \\ 
    \log \,\pi_{K1}^{(t)}(0, \bm{0}) & \log \,\pi_{K2}^{(t)}(0,\bm{0}) & \dots  & \log \,\pi_{KK}^{(t)}(0,\bm{0})
    \end{bmatrix},
\end{align*}
it follows that 
\begin{align}
\label{eq:matrix}
    \bm{A_{0}}\left(\bm{\Xi}^{(t)}\right)  \left(\left(\bm{\Pi_{0}}\left(\bm{\Xi}^{(t)}\right)\right)^{\top}+ \bm{\Pi_{0}}\left(\bm{\Xi}^{(t)}\right) \right) = \left(\Omega_{ik}\left(\bm{0},\bm{0}, \bm{\Xi}^{(t)}\right)\right)_{ik}
\end{align}
holds. Put differently, we are able to write $\Omega_{ik}^{(t)}(\bm{0},\bm{0}, \bm{\Xi})$ as the $(i,k)$th entry of the result of \eqref{eq:matrix}. 
To compute $\bm{\Pi_{0}}\left(\bm{\Xi}^{(t)}\right)$, we first evaluate the matrix $\pi^{(t)}(1, \bm{0}) \in [0,1]^{K\,\times\, K}$ via sparse matrix operations:
\begin{align}
    \label{eq:p_0}
    \bm{\pi}^{(t+1)}(1,\bm{0}) = \left(\left(\bm{\Xi}^{(t+1)}\right)^{\top} \bm{g} \circ (\bm{J} - \bm{X_{1}}) \circ \cdots \circ (\bm{J} - \bm{X_{p}})\ \left(\bm{\Xi}^{(t+1)}\right) \right) \oslash \\\notag
    \left(\left(\bm{\Xi}^{(t+1)} \right)^{\top} (\bm{J} - \bm{X_{1}}) \circ \cdots \circ (\bm{J} - \bm{X_{p}}) \left(\bm{\Xi}^{(t+1)} \right)\right),
\end{align}
where $\bm{A} \circ \bm{B}$ denotes the Hadamard (i.e., entry-wise) product of the conformable matrices $\bm{A}$ and $\bm{B}$, $\bm{J}$ is a $N \times N$ matrix whose off-diagonal entries are all one and whose diagonals are all zero. 
We follow the approach \citet{DahburaEtAl2021} to calculate \eqref{eq:p_0} without breaking the sparsity of the covariate matrices and $\bm{g}$. We can then set 
\begin{align*}
    \bm{\Pi_{0}}\left(\bm{\Xi}^{(t)}\right) = \log \left(1- \bm{\pi}^{(t+1)}(1,\bm{0})\right)
\end{align*}
and calculate $\Omega_{ik}^{(t)}\left(\bm{0}, \bm{0}, \bm{\Xi}^{(t)}\right)$ for $i = 1, \ldots, N$ and $k = 1, \ldots, K$. 

Next, we correct for the error arising from assuming that $\bm{g} = \bm{0}$ and $\bm{x} = \bm{0}$ holds. Since all covariates are assumed to be categorical, the pairwise covariate vector $\bm{x}_{ij}$  can have at most $2^p$ possible outcomes, which we collect in the set $\mathcal{X}$.  
Given this, one may rewrite the sum over $j \neq i$ as a sum over all possible outcomes of the pairwise covariate information: 
\begin{align*}
   \Lambda_{ik}\left(\bm{g}, \bm{x}, \bm{\Xi}^{(t)}\right)~&=~         \sum_{j \neq i}^{N}  \sum_{l=1}^{K} \left(g_{ij} \xi_{jl}^{(t)} \left(\log \,\dfrac{\pi_{kl}^{(t)} (1, \bm{x}_{ij})}{\pi_{kl}^{(t)} (0, \bm{0})}\right) + g_{ji} \xi_{jl}^{(t)} \left(\log \,\dfrac{\pi_{lk}^{(t)} (1, \bm{x}_{ji})}{\pi_{lk}^{(t)} (0, \bm{0})}\right)\right) \\
   &+\, \sum_{j \neq i}^{N}  \sum_{l=1}^{K} \left((1-g_{ij}) \xi_{jl}^{(t)} \left(\log \,\dfrac{\pi_{kl}^{(t)} (0, \bm{x}_{ij})}{\pi_{kl}^{(t)} (0, \bm{0})}\right) + (1-g_{ji}) \xi_{jl}^{(t)} \left(\log \,\dfrac{\pi_{lk}^{(t)} (0, \bm{x}_{ji})}{\pi_{lk}^{(t)} (0, \bm{0})}\right)\right)\\ 
    &= \sum_{\bm{x} \in \mathcal{X}} \sum_{j \neq i}^{N}  \sum_{l=1}^{K}g_{ij}\, \mathbb{I}(\bm{x}_{ij} = \bm{x}) \xi_{jl}^{(t)} \left(\log \,\dfrac{\pi_{kl}^{(t)} (1, \bm{x})}{\pi_{kl}^{(t)} (0, \bm{0})}\right)\\
    &+\,g_{ji}\, \mathbb{I}(\bm{x}_{ji} = \bm{x}) \xi_{jl}^{(t)} \left(\log \,\dfrac{\pi_{lk}^{(t)} (1, \bm{x})}{\pi_{lk}^{(t)} (0, \bm{0})}\right) + (1-g_{ij})\, \mathbb{I}(\bm{x}_{ij} = \bm{x}) \, \xi_{jl}^{(t)} \left(\log \,\dfrac{\pi_{kl}^{(t)} (0, \bm{x})}{\pi_{kl}^{(t)} (0, \bm{0})}\right)\\ 
    &+\,1-g_{ji})\, \mathbb{I}(\bm{x}_{ji} = \bm{x})\,  \xi_{jl}^{(t)} \left(\log \,\dfrac{\pi_{lk}^{(t)} (0, \bm{x})}{\pi_{lk}^{(t)} (0, \bm{0})}\right)
\end{align*}
With 
\begin{align*}
    \bm{\Pi}\left(d,\bm{x}, \bm{\Xi}^{(t)}\right) \coloneqq 
    \begin{bmatrix} 
    \log \frac{\pi_{11}^{(t)}(d, \bm{x})}{\pi_{11}^{(t)}(d, \bm{0})} &    \log \frac{\pi_{12}^{(t)}(d, \bm{x})}{\pi_{12}^{(t)}(d, \bm{0})} & \dots &   \log \frac{\pi_{1K}^{(t)}(d, \bm{x})}{\pi_{1K}^{(t)}(d, \bm{0})} \\
      \log \frac{\pi_{21}^{(t)}(d, \bm{x})}{\pi_{21}^{(t)}(d, \bm{0})}  &  \log \frac{\pi_{22}^{(t)}(d, \bm{x})}{\pi_{22}^{(t)}(d, \bm{0})}  & \dots &   \log \frac{\pi_{2K}^{(t)}(d, \bm{x})}{\pi_{2K}^{(t)}(d, \bm{0})} \\
    \vdots & \vdots  & \ddots  & \vdots \\ 
  \log \frac{\pi_{K1}^{(t)}(d, \bm{x})}{\pi_{K1}^{(t)}(d, \bm{0})} &   \log \frac{\pi_{K2}^{(t)}(d, \bm{x})}{\pi_{K2}^{(t)}(d, \bm{0})}  & \dots  &   \log \frac{\pi_{KK}^{(t)}(d, \bm{x})}{\pi_{KK}^{(t)}(d, \bm{0})} 
    \end{bmatrix},
\end{align*}
\begin{align*}
    \bm{\Gamma}(d) \coloneqq 
    \begin{cases}
    \bm{g} & (d = 1) \\
    \bm{J} - \bm{g} & (d = 0)
    \end{cases},
\end{align*}
and
\begin{align*}
    \bm{\Delta}(\bm{x}) \coloneqq \bm{X_{1}}^{x_1}\circ (\bm{J} - \bm{X_{1}})^{1-x_1} \circ \cdots \circ \bm{X_{p}}^{x_p}\circ (\bm{J} - \bm{X_{p}})^{1-x_p},
\end{align*}
we can write 
\begin{align*}
    \Lambda_{ik}&\left(\bm{g}, \bm{x}, \bm{\Xi}^{(t)}\right) =   
    &\left(\sum_{\bm{x} \in \mathcal{X}} \sum_{d \in \{0,1\}}  \bm{\Gamma}(d) \circ \bm{\Delta}(\bm{x}) \,  \bm{\Xi}^{(t)} \, \bm{\Pi}^{(t)}\left(d,\bm{x}, \bm{\Xi}^{(t)}\right) + \left(\bm{\Gamma}(d) \circ \bm{\Delta}(\bm{x})\right)^\top\,  \bm{\Xi}^{(t)} \, \bm{\Pi}^{(t)}\left(d,\bm{x}, \bm{\Xi}^{(t)}\right)^\top\right)_{ik} .
\end{align*}
Extending \eqref{eq:p_0} to generic covariate values $\bm{x}$ yields 
\begin{align*}
    \bm{\pi}^{(t+1)}(1, \bm{x}) = &\big(\left(\bm{\Xi}^{(t+1)}\right)^{\top} \bm{g} \circ \bm{X_{1}}^{x_1}\circ (\bm{J} - \bm{X_{1}})^{1-x_1} \circ \cdots \circ \bm{X_{p}}^{x_p}\circ (\bm{J} - \bm{X_{p}})^{1-x_p} \left(\bm{\Xi}^{(t+1)}\right) \big) \oslash \\ 
    &\left(\left(\bm{\Xi}^{(t+1)} \right)^{\top} \bm{X_{1}}^{x_1}\circ (\bm{J} - \bm{X_{1}})^{1-x_1} \circ \cdots \circ \bm{X_{p}}^{x_p}\circ (\bm{J} - \bm{X_{p}})^{1-x_p} \left(\bm{\Xi}^{(t+1)}\right)\right),
\end{align*}
enabling the evaluation of $\bm{\Pi}\left(d,\bm{x}, \bm{\Xi}^{(t)}\right)$ for $\bm{x} \in \mathcal{X}$.
Following  \citet{DahburaEtAl2021}, we can still evaluate this matrix by sparse matrix operations. 

In sum, we have shown how to compute the terms $ \Omega_{ik}\left(\bm{0}, \bm{0}, \bm{\Xi}^{(t)}\right)$ and $\Lambda_{ik}\left(\bm{g}, \bm{x}, \bm{\Xi}^{(t)}\right)$ through matrix multiplications. 
Plugging in these terms into \eqref{eq:decomposition} enables a fast computation of $\Omega_{ik}(\bm{g}, \bm{x}, \bm{\Xi}^{(t)})$, which is the computational bottleneck for evaluating the quadratic term needed to update $\bm{\Xi}^{(t)}$ to $\bm{\Xi}^{(t+1)}$. 

\section{Additional figures and estimates: Comparison with other models}\label{Sec::Appendix:D}

\begin{figure}
    \centering
    \includegraphics[width=\linewidth]{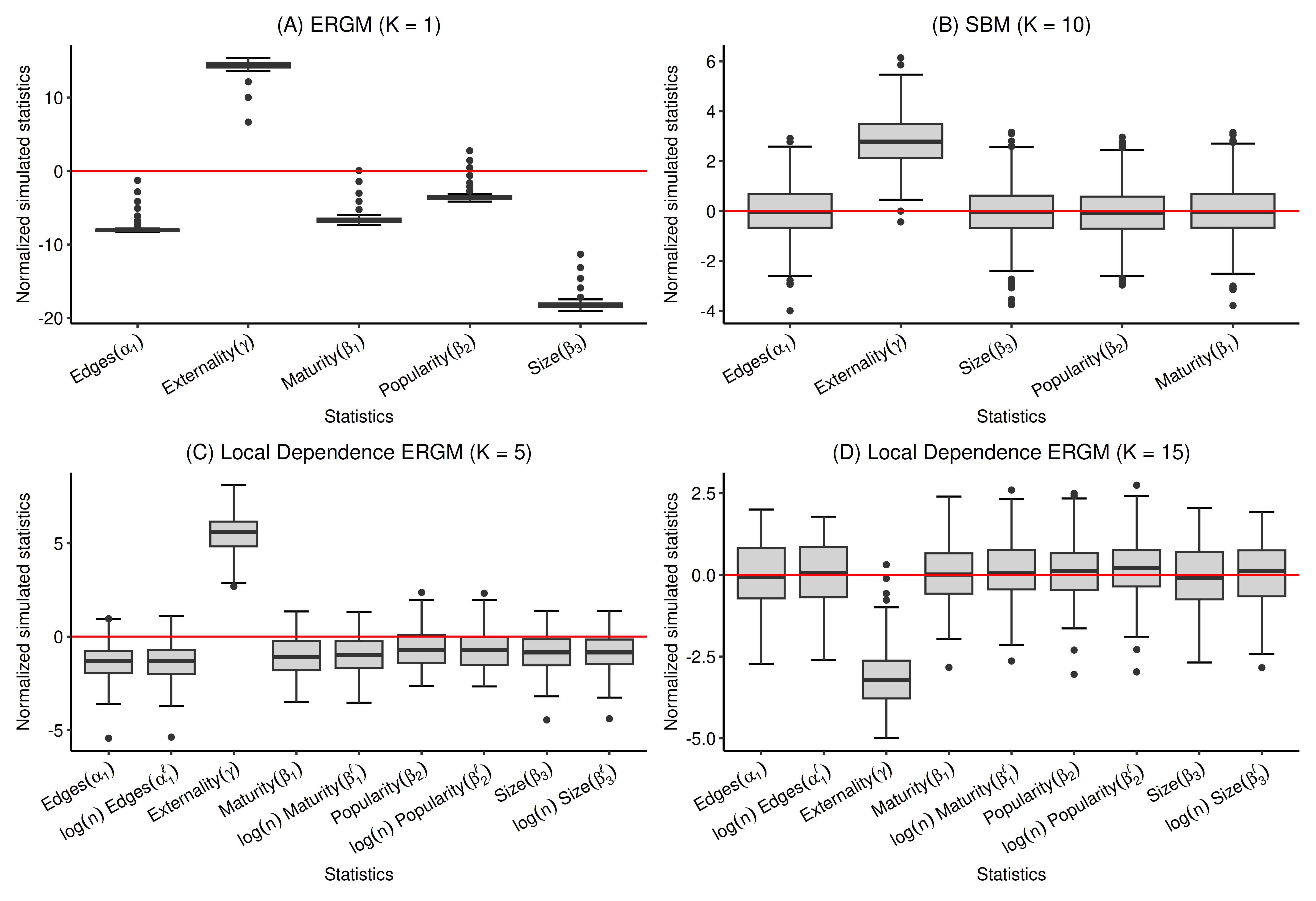}
    \caption{Comparison of model fits of network statistics of competing models. }
    \label{fig:gof_appendix}
    \flushleft \scriptsize Each boxplot is based on 100 simulations of the network generated using the estimated parameters from the corresponding model. 
    Simulations are initialized at the observed network and proceed with a burn-in period of \mbox{500,000} Metropolis–Hastings steps. 
    After burn-in, one network is retained every 50,000 steps. For each retained network, the corresponding network statistics are computed and then standardized such that a value of zero corresponds to the statistic of the observed network. 
    On the x-axis, one boxplot is displayed for each network statistic associated with a coefficient the corresponding fitted model.
\end{figure}

In Figure \ref{fig:gof_appendix} (A), we show the simulated statistics generated by a model with $K=1$, corresponding to an Exponential Random Graph Model (ERGM). 
In aggregate, this model performs very poorly, which suggests that the unobserved heterogeneity helps the model fit. 

We compare our model with $K=10$ unobserved types to a stochastic blockmodel with $K=10$ unobserved types in Figure \ref{fig:gof_appendix} (B). 
In essence, the comparison model is the same as our model without the externality. 
As expected, the stochastic blockmodel (SBM) performs worse in predicting the externality within blocks.


We also performed robustness checks to choose the number of types $K$. 
In Figures \ref{fig:gof_appendix} (C) and (D), we report the boxplot for the fit of models with $K=5$ and $K=15$, respectively. 

We conclude from the plots that increasing or decreasing the number of types does not seem to help in representing the externality. 
We report the estimates of all these competing models in Table \ref{tbl:results_python_suppl}.

\begin{table}[!h]
\caption{Parameter estimates and standard errors.}\label{tbl:results_python_suppl} 
\begin{center}
\begin{tabular}{llrcrr}
\toprule
\multicolumn{1}{l}{\  }&\multicolumn{2}{c}{\  ERGM (K = 1)}&\multicolumn{1}{l}{\  }&\multicolumn{2}{c}{\   SBM (K = 10)}\tabularnewline
\cline{2-3} \cline{5-6}
&Within&Between&&Within&Between\tabularnewline
&Estimate&Estimate&&Estimate&Estimate\tabularnewline
&(Std. Error)&(Std. Error)&&(Std. Error)&(Std. Error)\tabularnewline
Edges ($\alpha_1$)&$-$7.961&&&$-$5.703&-7.979\tabularnewline
 &(0.007)&&&(0.274)&(0.006)\tabularnewline
$\log(n)\times$ Edges ($\alpha_2)$&&&&$-$0.549&\tabularnewline
 &&&&(0.031)&\tabularnewline
Externality ($\gamma$)&$-$0.178&&&&\tabularnewline
 &(0.002)&&&&\tabularnewline
Maturity ($\beta_1$)&$-$0.131&&&1.767&-0.222\tabularnewline
 &(0.01)&&&(0.303)&(0.011)\tabularnewline
$\log(n) \times$ Maturity \big($\beta_1^\ell$\big)&&&&$-$0.12&\tabularnewline
 &&&&(0.035)&\tabularnewline
Popularity ($\beta_2$)&$-$0.066&&&1.071&-0.07\tabularnewline
 &(0.01)&&&(0.308)&(0.011)\tabularnewline
$\log(n)\times$Popularity \big($\beta_2^\ell$\big)&&&&$-$0.117&\tabularnewline
 &&&&(0.035)&\tabularnewline
Size ($\beta_3$)&0.127&&&1.92&0.118\tabularnewline
 &(0.01)&&&(0.303)&(0.01)\tabularnewline
$\log(n)\times$Size \big($\beta_3^\ell$\big)&&&&$-$0.178&\tabularnewline
 &&&&(0.035)&\tabularnewline
 \midrule
\end{tabular}
\end{center}
\end{table}

\clearpage

\begin{table}[ht]
\begin{center}
Table A1 (ctd.)
\begin{tabular}{llrcrr}
\midrule
\multicolumn{1}{l}{\  }&\multicolumn{2}{c}{\  Local Dep. ERGM (K = 5)}&\multicolumn{1}{l}{\  }&\multicolumn{2}{c}{\  Local Dep. ERGM (K = 15)}\tabularnewline
\cline{2-3} \cline{5-6}
&Within&Between&&Within&Between\tabularnewline
&Estimate&Estimate&&Estimate&Estimate\tabularnewline
&(Std. Error)&(Std. Error)&&(Std. Error)&(Std. Error)\tabularnewline
Edges ($\alpha_1$)&$-$4.893&-7.45&&$-$4.181&-8.28\tabularnewline
 &(0.263)&(0.006)&&(0.305)&(0.006)\tabularnewline
$\log(n)\times$ Edges ($\alpha_2)$&$-$0.569&&&$-$0.765&\tabularnewline
 &(0.028)&&&(0.038)&\tabularnewline
Externality ($\gamma$)&$-$0.258&&&0.135&\tabularnewline
 &(0.013)&&&(0.013)&\tabularnewline
Maturity ($\beta_1$)&0.711&-0.267&&3.201&-0.198\tabularnewline
 &(0.302)&(0.012)&&(0.321)&(0.011)\tabularnewline
$\log(n) \times$ Maturity \big($\beta_1^\ell$\big)&$-$0.012&&&$-$0.307&\tabularnewline
 &(0.033)&&&(0.041)&\tabularnewline
Popularity ($\beta_2$)&1.324&-0.073&&0.369&-0.074\tabularnewline
 &(0.307)&(0.011)&&(0.29)&(0.01)\tabularnewline
$\log(n)\times$Popularity \big($\beta_2^\ell$\big)&$-$0.142&&&$-$0.048&\tabularnewline
 &(0.033)&&&(0.037)&\tabularnewline
Size ($\beta_3$)&1.316&0.104&&1.58&0.117\tabularnewline
 &(0.31)&(0.011)&&(0.287)&(0.01)\tabularnewline
$\log(n)\times$Size \big($\beta_3^\ell$\big)&$-$0.111&&&$-$0.15&\tabularnewline
 &(0.034)&&&(0.037)&\tabularnewline
\hline
\end{tabular}
\end{center}
\flushleft Estimates and standard errors are obtained by Maximum Pseudolikelihood (MPLE), conditioning on the estimated types in the first step. Standard error for the estimates are in parenthesis. 
\end{table}

 \clearpage
%

\end{document}